\documentclass[aps,pra,reprint,superscriptaddress,nofootinbib]{revtex4-1}

\usepackage{graphicx} 
\usepackage{amssymb}
\usepackage{amsmath}
\usepackage{amsthm}   
\usepackage{bbm}      
\usepackage{tikz}
\usepackage[pdfencoding=auto, psdextra]{hyperref} 

\newcommand{\keyword}{\emph}

\newcommand{\commute}[2]{\left[#1,#2\right]}
\newcommand{\acommute}[2]{\left\{#1,#2\right\}}

\newcommand{\bbra}[1]{\left\langle\kern-.5ex\middle\langle #1 \right\rvert}
\newcommand{\kket}[1]{\left\lvert #1 \middle\rangle\kern-.5ex\right\rangle}
\newcommand{\bbrakket}[2]{\left\langle\kern-.5ex\middle\langle #1 \middle\vert #2 \middle\rangle\kern-.5ex\right\rangle}
\newcommand{\bbraopkket}[3]{\left\langle\kern-.5ex\middle\langle #1 \middle\vert #2 \middle\vert #3 \middle\rangle\kern-.5ex\right\rangle}
\newcommand{\kketbbra}[2]{\left\lvert #1 \middle\rangle\kern-.5ex\middle\rangle \middle\langle\kern-.5ex\middle\langle #2 \right\rvert}

\newcommand{\probof}[1]{\textrm{P}\kern-.3ex\left( #1 \right)}
\newcommand{\probgiven}[2]{\textrm{P}\kern-.3ex\left( #1 \middle\vert #2 \right)}
\newcommand{\expect}[1]{\mathbb{E}\kern-.3ex\left[ #1 \right]}
\newcommand{\kldiv}[2]{\mathcal{D}_{\textrm{KL}}\kern-.5ex\left( #1 \middle\Vert #2 \right)}

\newcommand{\trans}{\textrm{T}}  
\newcommand{\draz}{+}            
\newcommand{\ident}{\mathbbm{1}} 
\newcommand{\onevec}{\boldsymbol{1}} 
\DeclareMathOperator{\pdet}{\overline{\det}} 
\DeclareMathOperator{\tr}{Tr} 

\newcommand{\numMeas}{N}  
\newcommand{\chainLen}{M} 
\newcommand{\seqLen}{L} 
\newcommand{\markOrd}{m} 

\graphicspath{{./figures/}}

\begin{document}

\title{The role of correlations in a sequence of quantum observations on empirical measures}
\date{November 12, 2024}
\author{Joseph A. Smiga}
\email{joseph.smiga@rochester.edu}
\affiliation{Department of Physics and Astronomy, University of Rochester, Rochester, NY 14627, USA}
\affiliation{Center for Coherence and Quantum Optics, University of Rochester, Rochester, NY 14627, USA}
\author{Gabriel T. Landi}
\email{glandi@ur.rochester.edu}
\affiliation{Department of Physics and Astronomy, University of Rochester, Rochester, NY 14627, USA}
\affiliation{Center for Coherence and Quantum Optics, University of Rochester, Rochester, NY 14627, USA}

\begin{abstract}
    The outcome of continuously measuring a quantum system is a string of data whose intricate correlation properties reflect the underlying quantum dynamics. 
    In this paper we study the role of these correlation in reconstructing the probabilities of finite sequences of outcomes, the so-called empirical distributions. 
    Our approach is cast in terms of generic quantum instruments, and therefore encompass all types of sequential and continuous quantum measurements. We also show how this specializes to important cases, such as quantum jumps.
    To quantify the precise role of correlations, we introduce a relative-entropy based measure that quantifies the range of correlations in the string, and the influence that these correlations have in reconstructing finite sequences. 
\end{abstract}

\maketitle

\section{Introduction}

When measuring a quantum system, the standard technique is the so-called ``\keyword{prepare and measure}'' approach, wherein the state of a system is reset after each measurement. 
However, this can be somewhat restrictive, and a broader approach would consist of sequentially measuring the same system over and over; i.e., the experiment is not reset between observations.
This brings about additional complexities because of the interplay between information gain and measurement backaction which now has to be explicitly taken into account.
The correlations between these measurement outcomes reveal crucial properties about the underlying quantum system, with famous examples including Glauber's coherence functions~\cite{plenio_quantum-jump_1998} or the shot noise in mesoscopic physics~\cite{buttiker1992}. 

A common class of sequentially measured systems are quantum jumps~\cite{plenio_quantum-jump_1998, wiseman2009, landi_current_2024}.
Although quantum mechanics often involves ``smooth'' unitary evolution, the introduction of quantum jumps can provide a powerful tool for describing sudden, irreversible processes that one may find in nature~\cite{piilo_non-markovian_2008, piilo_open_2009, landi_patterns_2023, hatridge_quantum_2013, blok_manipulating_2014}; such as spontaneous transitions, measurements, and interactions with the environment. 
Correlations between quantum jumps can be highly non-trivial and play a significant role in the underlying stochastic dynamics~\cite{aharonov_fault-tolerant_2006, zou_spatially_2024, bagewadi_effect_2024, schmid_repeated_1987, dittrich_continuous_1990, giovannetti_quantum_2006, yang_memory_2019, altherr_quantum_2021}. 
Much focus has been placed on finding ways to minimize these correlations, such as through quantum non-demolition measurements~\cite{caves_measurement_1980}, to simplify the dynamics of the system.
However, correlations do not inherently improve or worsen the results of the measurements~\cite{radaelli_fisher_2023, zou_spatially_2024}. Thus, further understanding the role correlations play, when not suppressed, can lead to improved developments in quantum technology.

In this paper we consider the task of reconstructing finite sequences from a string of measurements of a quantum systems. 
The basic problem is illustrated in Fig.~\ref{fig:seq_empDists}. 
A continuously measured quantum system yields a very long string of data, which is generally highly correlated. 
From these data one wishes to reconstruct the probabilities of finite sequences. These are the so-called \emph{empirical distributions} (EDs) and represent a form of data compression. 
The simplest one, which we denote ED1, simply consists of counting how many times we obtain each outcome, in this case 0 or 1 (left histogram in Fig.~\ref{fig:seq_empDists}). 
Because the data can be correlated, it is also of interest to compute higher order statistics: ED2, for instance, counts how many times each of the four possible sequences, $\{00, 01, 10, 11\}$, appear in the string (middle histogram in Fig.~\ref{fig:seq_empDists}). ED3 would count triplets, like 001 and 010 (right histogram in Fig.~\ref{fig:seq_empDists}), and so on and so forth for higher order empirical distributions. 

Summary statistics such as the multi-point correlation function~\cite{tilloy_exact_2018, guilmin_correlation_2023, guilmin_parameters_2024} and the empirical distribution explored here are capable of compressing large data sets while retaining some useful information around complex patterns. 
Motivated by this, we provide two main results. 
First, we provide general methods to calculate the covariance matrix between entries of the empirical distribution of arbitrary order. 
This contains information on how correlations in the outcomes affect our ability to reconstruct finite empirical distributions. 
These formulas are useful within the context of full counting statistics, since they are the basic building blocks used in assessing current fluctuations~\cite{landi_current_2024}, a topic that has recently been the subject of intense interest in the context of thermo-kinetic uncertainty relations~\cite{vo_unified_2022, hasegawa_unifying_2023,hasegawa2019,guarnieri2019,hasegawa2020,hasegawa2021,timpanaro2019a,prech2023a}.  
As our second result, we introduce an information-theoretic measure, based on the relative entropy, which gauges how correlations in the empirical distributions affect our ability to reconstruct the process under simple (Markovian) assumptions. 
This yields insight into the role of complex correlations between strings of data. 

Our formalism is based on general quantum instruments, which therefore encompass the most general type of sequential quantum measurement~\cite{wiseman2009, milz_quantum_2021}.
In addition to being more general, we also show that this greatly simplifies the mathematical framework.
Connections with specific measurement schemes, particularly quantum jumps, are also discussed in detail. 
The work here complements previous work~\cite{radaelli_fisher_2023, radaelli_parameter_2024} that focused on the role of correlations in metrology and parameter estimation. 
Here our focus is on the correlations themselves. 
Moreover, previous work focused almost exclusively on ED1, while here our focus is a general ED$\seqLen$.

The paper is organized as follows: Sec.~\ref{sec:important_concepts} summarizes important concepts for the paper, Sec.~\ref{sec:empCorrInfo_quant} presents a key measure of correlations for the empirical distribution of a quantum process (using calculations in Appendix~\ref{app:PsiMat}), an information theoretic measure of the role correlations play is given in Sec.~\ref{sec:relEnt_empDist}, and illustrative examples are given in Sec.~\ref{sec:examples}. Finally, concluding remarks are given in Sec.~\ref{sec:conclusion}.

\section{Important concepts}\label{sec:important_concepts}
To start, we describe important background concepts in this section. 
We consider systems that are sequentially measured yielding a string like that illustrated in Fig.~\ref{fig:seq_empDists}. 
The data have the form $\{X_1,X_2,\ldots,X_{\numMeas} \}$ and we assume that each symbol $X_i$ is always drawn from a finite alphabet $\mathbb{M}$ (the ``observable symbols'').
The underlying distribution is written as $\probof{X_1,X_2,\ldots}$ and can be arbitrarily complicated. 
Our focus will always be on scenarios involving a large number of data points, $\numMeas\gg1$. 
We therefore focus on stationary conditions, in which the statistics of substrings are translationally invariant. 
This means, for example, that marginal distributions such as $\probof{X_i}$ (obtained by summing over all outcomes except $X_i$) are assumed to be independent of $i$. The same is true for contiguous blocks, such as $\probof{X_i,X_{i+1}}$, and so on.

\subsection{Empirical distributions}\label{sec:bkg_empDist}
A common method for ``compressing'' a long sequence $\{X_1,X_2,\ldots,X_{\numMeas} \}$ is to generate a histogram of outcomes; i.e., to count how many times each symbol appears. 
This is called the \keyword{empirical distribution}~\cite{bertini_large_2015, smiga_stochastic_2023, stewart_consistency_2024}.
Because the data are correlated, however, to fully characterize the statistics we are interested in counting how many times we observe higher order substrings, such as pairs of symbols, triplets, and so on.  
This generates a family of empirical distributions, which we refer to as ED$\seqLen$, where $\seqLen$ is the size of the substring we are counting. 
The simplest case is ED1, which corresponds to just counting how many times each symbol appears:
\begin{equation}\label{eq:empdist_general}
    q_x = \frac{1}{\numMeas} \sum_{i=1}^{\numMeas} \delta_{X_i,x}\,,
\end{equation}
for $x\in\mathbb{M}$.
The ED1 is  normalized as $\sum_x q_x = 1$.
If the process is stationary, the ED1 will be an unbiased estimator of the marginal distribution 
of individual symbols~\cite{smiga_stochastic_2023}. That is, $\expect{q_x} = \probof{X}$, where $\expect{\bullet}$ means an average over ensembles.

\begin{figure}[ht]
    \centering
    \includegraphics[width=\linewidth]{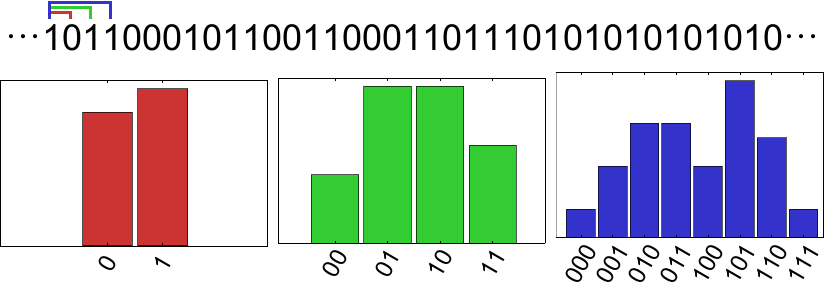}
    \caption{Reconstructing the empirical distributions from a string of data. Above is a portion of the string, while below are empirical distributions for different lengths of substrings. 
    The distribution ED$\seqLen$ (here illustrated for $\seqLen=1,2,3$ colored red, green, and blue, respectively) corresponds to counting how many times each substring of length $\seqLen$ occurs as we parse along the string. For example, ED2 counts how many times we observe 00, 01, 10, and 11. 
    }
    \label{fig:seq_empDists}
\end{figure}

In a similar fashion one can also define empirical distributions for substrings of arbitrary length. 
For example, the empirical distribution of pairs of outcomes, ED2, is defined as
\begin{equation}\label{eq:ED2}
    q_{xy} = \frac{1}{\numMeas-1}\sum_{j=1}^{\numMeas-1} \delta_{X_j,x}\delta_{X_{j+1},y}.
\end{equation}
It counts how many times a pair of $(x,y)$ appears in a sequence (c.f. the middle histogram in Fig.~\ref{fig:seq_empDists}). 
As before, ED2 is also a unbiased estimator for the true joint distribution of pairs, $\expect{q_{x,y}} = \probof{X,Y}$.
And it is normalized as $\sum_{x,y} q_{xy} = 1$.
Proceeding in a similar fashion, the expression for ED$\seqLen$ reads 
\begin{equation}\label{eq:empdist_seq_general}
    q_{\boldsymbol{x}} = \frac{1}{\numMeas-\seqLen+1} \sum_{j=1}^{\numMeas-\seqLen+1} \delta_{x_1,X_j}\delta_{x_2,X_{j+1}}\cdots \delta_{x_{\seqLen},X_{j+\seqLen-1}}\,,
\end{equation}
where $\boldsymbol{x} = (x_1,\ldots,x_\seqLen)$.

Marginalizing the ED$\seqLen$ $q_{\boldsymbol{x}}$ over one if its indices yields ED($\seqLen-1$). 
However, there exists additional constraints on the empirical distributions, which at first might not be very apparent. 
For example, if we marginalize ED2 over $x$ or over $y$ we should recover the same ED1 distribution: 
\begin{equation}
    \sum_y q_{xy} = \sum_y q_{yx} = q_x.
\end{equation}
If $d = \lvert\mathbb{M}\rvert$ is the total number of symbols, then ED$\seqLen$ will have a total of $d^{\seqLen-1}$ linear constraints of this form (including normalization).
As we discuss in Appendix~\ref{app:seq_constraints}, this is related to a the fact that for ED$\seqLen$ symbols in a string are always counted $\seqLen-1$ times. 
For example, if we estimate ED2 in a string like
\[
    \bullet\bullet\bullet\bullet 101\bullet\bullet\bullet\bullet
\]
the outcome 0 will enter in the count of $q_{10}$, as well as in the count of $q_{01}$.

\subsection{The covariance matrix of the empirical distribution}

The empirical distribution is a function of the random outcomes, and is therefore a random variable itself. 
The statistics of the empirical distribution follow from the statistics of the underlying data and, because of correlations between outcomes, its distribution can be quite complicated.
In Ref.~\cite{smiga_stochastic_2023}, these complexities were avoided by considering the limit of a large number of outcomes, and approximating the distribution $\probof{\boldsymbol{q}}$ as a multivariate Gaussian distribution\footnote{The validity of this approximation is assumed in Ref.~\cite{smiga_stochastic_2023}. Roughly, this follows from arguments similar to the central limit theorem as long as correlations decay sufficiently quickly. An example of a stationary process in which this assumption does not hold is described in Sec.~\ref{sec:periodic_chain}.} whose dimension is the cardinality of the alphabet $\lVert\mathbb{M}\rVert$. 
With this assumption, the mean and covariance matrix fully determine the empirical distributions. 
Since, as argued above, the EDs are unbiased estimators for the respective substring probabilities, the mean of ED1, ED2, etc. will be given precisely by $\probof{X_i}$, $\probof{X_i, X_{i+1}}$, etc. 
In this section we discuss how to calculate the second moments.

We start with ED1. 
The covariance matrix is defined as 
\begin{equation}
    \Sigma_{xy} :=\expect{q_x q_y} - \expect{q_x}\expect{q_y},
\end{equation}
and depends on the correlations between measurements. From the full probability of the data, $\probof{X_1,\ldots,X_\numMeas}$, define the conditional probability of measuring $y$ after $\ell$ steps from measuring $x$:
\begin{equation}\label{eq:cond_observation}
    p_{y\gets  x}(\ell) = \probgiven{X_{i+\ell}=y}{X_i=x}\,,
\end{equation}
which, assuming a stationary process, is independent of $i$. Then define the matrix
\begin{equation}\label{eq:PsiMat}
    \Psi_{yx} = \sum_{\ell=1}^{\numMeas-1} \left( 1-\frac{\ell}{\numMeas} \right)\left[ p_{y\gets  x}(\ell)-p_y \right]\,,
\end{equation}
for $p_y=\probof{X_i=y}$. For uncorrelated measurements and $\ell\geq 1$, one finds $p_{y\gets  x}(\ell)=p_y$ and $\Psi$ vanishes. Thus $\Psi$ can be understood as encoding correlations between measurements. 
As shown in Ref.~\cite{smiga_stochastic_2023}, the covariance matrix  is then given by 
\begin{equation}\label{eq:empdist_covar}
\begin{aligned}
    \Sigma_{xy} &= \frac{1}{\numMeas}\left[ p_x \left( \delta_{xy} - p_y \right) + \Psi_{yx}p_x + \Psi_{xy}p_y \right]\,.
\end{aligned}    
\end{equation}
The covariance matrix decays with the number of data points $N$. 
For uncorrelated data $\Psi=0$ and the covariance is given only by the first term. 
Notice also that, because $\sum_x q_x = 1$ the entries of the empirical distribution are actually perfectly correlated. 
That is $\det\Sigma =0$. In practice, of course, thus just means that it suffices to consider $\lVert\mathbb{M}\rVert-1$ symbols $x$ to fully characterize $q_x$.

The calculation of the covariance matrix for ED$\seqLen$ is analogous, with the main difference being that $\Psi$ and $\boldsymbol{p}$ are indexed by finite sequences of measurements $\boldsymbol{x} = (x_1,\ldots,x_\seqLen)$ instead of individual measurements. 
The formulas read 
\begin{equation}\label{eq:PsiMat_seq}
    \Psi_{\boldsymbol{y}\boldsymbol{x}} = \sum_{\ell=1}^{\numMeas-\seqLen} \left( 1-\frac{\ell}{\numMeas-\seqLen+1} \right)\left[ p_{\boldsymbol{y}\gets \boldsymbol{x}}(\ell)-p_{\boldsymbol{y}} \right]\,,
\end{equation}
for
\begin{equation}\label{eq:cond_observation_seq}
    p_{\boldsymbol{y}\gets \boldsymbol{x}}(\ell) = \probgiven{X_{i+\ell:i+\ell+\seqLen}=\boldsymbol{y}}{X_{i:i+\seqLen}=\boldsymbol{x}}\,.
\end{equation}
Here we used the shorthand $X_{a:b}=\{X_a,X_{a+1},\ldots,X_b\}$. The above equation describes the probability of measuring the sequence $\boldsymbol{y}$ after $\ell$ steps from measuring sequence $\boldsymbol{x}$ (counting from the start of the sequences).
The covariance matrix will then have the exact same structure
\begin{equation}
    \Sigma_{\boldsymbol{x} \boldsymbol{y}} 
    = \frac{1}{\numMeas-\seqLen+1}
    \left[ 
    p_{\boldsymbol{x}}
    \left( 
    \delta_{\boldsymbol{x}\boldsymbol{y}} - p_{\boldsymbol{y}} 
    \right) + 
    \Psi_{\boldsymbol{y}\boldsymbol{x}}p_{\boldsymbol{x}} + \Psi_{\boldsymbol{x}\boldsymbol{y}}p_{\boldsymbol{y}} 
    \right],
\end{equation}
where $\delta_{\boldsymbol{x},\boldsymbol{y}} = \delta_{x_1,y_1}\ldots \delta_{x_\seqLen,y_\seqLen}$.
When computing Eq.~\eqref{eq:cond_observation_seq} some care must be taken with strings that partially overlap. 
For example, consider the case of ED2 and let $\boldsymbol{x} =(x_1,x_2)$ and $\boldsymbol{y} =(y_1,y_2)$.
Then $p_{\boldsymbol{y}\gets \boldsymbol{x}}(\ell)$ for $\ell = 1$ will vanish unless $x_2 = y_1$ since, by definition,
\begin{align*}
    p_{\boldsymbol{y}\gets \boldsymbol{x}}(1) &= \probgiven{X_{i+1} = y_1, X_{i+2}=y_2}{X_i = x_1, X_{i+1} = x_2}\\
    &= \delta_{y_1x_2} \probgiven{X_{i+2}=y_2}{X_i = x_1, X_{i+1} = x_2}\,.
\end{align*}
For $\ell > 1$ there will be no overlap to consider. 
Similarly, for ED3 one needs to consider overlaps for $\ell = 1$ and $\ell=2$, but not for $\ell >2$, and so on for ED$\seqLen$.

It is worth noting that all of the above results are absolutely general. 
They do not assume anything about the process that is generating them, which can be quantum or classical. 
They also do not assume Gaussianity.
To be able to estimate the covariance matrix, however, one needs to know the underlying probability $\probof{X_1,\ldots,X_\numMeas}$ and compute the sums in e.g. Eqs.~\eqref{eq:PsiMat_seq} and~\eqref{eq:cond_observation_seq}, which can be rather complicated.

\subsection{Sequential quantum measurements}\label{sec:quantum_instruments}

The effects of observing a quantum system on its dynamics can be generalized using \keyword{quantum instruments}~\cite{davies_operational_1970, lindblad_non-markovian_1979, milz_quantum_2021}. In this formalism, the evolution of the system is described by a set of trace-non-preserving superoperators indexed by the observation that, when acting on a density operator, describe the state of the system after making the respective observation. The trace of the outcome describes the probability of making the respective observation.

A common example arises from \keyword{positive operator-valued measures} (POVMs). A general POVM is described by a set of positive operators $\{F_x\}$ such that $\sum_x F_x=\ident$, where $x$ labels the outcome of a measurement. Define a corresponding set of operators\footnote{Strictly speaking, some care is needed to guarantee the existence of these operators. For the sake of this example, it is assumed that such a set of operators can be chosen.} $\{M_x\}$ such that $M_x^\dagger M_x = F_x$, and the superoperators $\{\mathcal{M}_x\}$ such that $\mathcal{M}_x\rho = M_x\rho M_x^\dagger$. These superoperators form the quantum instruments as the post-observation state $\rho\overset{x}{\mapsto}\rho'\propto\mathcal{M}_x\rho$ with $\probof{x}=\tr\mathcal{M}_x\rho$. One could also include the evolution of the system between observations by inserting unitary evolution operators in the instrument's superoperators.
Instruments therefore encompass not only the measurements, but also any kind of dynamical evolution in between.

More formally, a quantum instrument is described by a set of superoperators $\{\mathcal{M}_x\}$ indexed by the observation $x\in \mathbb{M}$. Given an initial state, $\rho_0$, the result of measuring $x_1$ is given by 
\begin{subequations}
\begin{align}
    \rho_0\overset{x_1}{\mapsto}\rho_1 &= \frac{\mathcal{M}_{x_1}\rho_0}{\probgiven{x_1}{\rho_0}} \\
    \textrm{for}\quad \probgiven{x_1}{\rho_0}&=\tr{\mathcal{M}_{x_1}\rho_0}\,,
\end{align}
\end{subequations}
where $\probgiven{x}{\rho_0}$ is also the probability of measuring $x$ given the initial state $\rho_0$. Note that throughout this work, this probability will be written without explicitly conditioning on the initial state (i.e., as $\probof{x}$) when the initial state is implied. 

Continuing, a second observation $x_2$ is sampled with probability,
\[
    \probgiven{x_2}{x_1} = \tr\big\{\mathcal{M}_{x_2}\rho_{x_1}\big\},
\]
and the state is updated to 
\[
    \rho_{x_{1:2}} = \frac{\mathcal{M}_{x_2}\rho_{x_1}}{\probgiven{x_2}{x_1}} = \frac{\mathcal{M}_{x_2}\mathcal{M}_{x_1}\pi}{\probof{x_{1:2}}},
\]
where $\probof{x_{1:2}} = \probgiven{x_2}{x_1} \probof{x_1}$.
Expanding on this to a string of measurements $\omega=x_{1:\numMeas}=\{x_1,\ldots,x_\numMeas\}$ results in
\begin{subequations}\label{eq:process_evolve}
    \begin{align}
        \rho_0\overset{\omega}{\mapsto} \rho_\omega &= \frac{\mathcal{M}_{x_\numMeas} \cdots \mathcal{M}_{x_1}\rho_0}{\probof{\omega}} \\
        \textrm{with}\quad \probof{\omega} &= \tr\left\{ \mathcal{M}_{x_\numMeas} \cdots \mathcal{M}_{x_1}\rho_0 \right\}\,.
    \end{align}
\end{subequations}

Further, one can define another superoperator, 
\begin{equation}\label{eq:superop_unknownResult}
    \mathcal{M}:=\sum_{x\in\mathbb{M}} \mathcal{M}_x\,,
\end{equation}
associated with making a measurement, but not knowing the result. This superoperator is trace-preserving: $\tr{\mathcal{M}\rho}=\tr{\rho}$. 
Using this superoperator, one can define the \keyword{steady state} of the system $\pi$ as one that satisfies 
\begin{equation}\label{eq:instrument_steady}
\mathcal{M}\pi=\pi\,. 
\end{equation}
This can be understood as the average state of a system immediately after a measurement is made. 
Under reasonable conditions, for any initial state $\rho$ we will have $\lim_{\ell\to\infty} \mathcal{M}^\ell\rho = \pi$.
We will henceforth take $\pi$ as the initial state since this automatically ensures that the distributions are stationary. Moreover, for long sequences  changing the initial state will lead to vanishingly small transient effects. 

Another useful statistic, which we will need, is the probability of a sequence of outcomes $x_{i_1}, \ldots, x_{i_\numMeas}$ which might not be contiguous. 
For example $\probof{x_1,x_4}$ or $\probof{x_1,x_3,x_5}$ and so on. 
This can be computed by simply  marginalizing Eq.~\eqref{eq:process_evolve} over the intermediate outcomes. Each symbol marginalized reduces to one power of the operator $\mathcal{M}$ in Eq.~\eqref{eq:superop_unknownResult}. Hence
\begin{align}\label{eq:jumpSeq_spaces}
    \probof{x_{i_1},\ldots,x_{i_{\numMeas}}} = \tr\{ & \mathcal{M}_{x_{i_\numMeas}} \mathcal{M}^{i_{\numMeas}-i_{\numMeas-1}-1} \cdots \nonumber\\
    & \cdots \mathcal{M}^{i_2-i_1-1}\mathcal{M}_{x_1}\mathcal{M}^{i_1-1} \rho\}\,,
\end{align}
with the proviso that $\mathcal{M}^0\to\ident$.

\subsection{Quantum jump process}
A special case of quantum instruments is the quantum jump processes~\cite{plenio_quantum-jump_1998, landi_patterns_2023}. This is described with a \keyword{Lindblad master equation} with the addition of dissipator superoperators,
\begin{equation}\label{eq:master_eq}
    \frac{d\rho}{dt} = \mathcal{L}\rho := -i\commute{H}{\rho} + \sum_x \mathcal{D}\left[L_x\right]\rho\,,
\end{equation}
$H$ is the Hamiltonian of the system, $\{L_x\}$ are the jump operators, and $\mathcal{D}[\bullet]$ is the \keyword{dissipator superoperator},
\begin{equation}
    \mathcal{D}\left[L\right]\rho := L\rho L^\dagger - \frac{1}{2} \acommute{L^\dagger L}{\rho}\,.
\end{equation}
The dissipator superoperators reflect different jumps that the system can undergo. However, not all of these jumps may be observable; for example, some jumps may reflect loss from an inefficient detector. Thus, one only considers the subset of observable jumps $x\in \mathbb{M}$.

To understand the state of a quantum system after measuring a quantum jump, it helps to first introduce the \keyword{jump superoperators} (c.f., Sec.~II.C in Ref.~\cite{landi_current_2024}),
\begin{equation}
    \mathcal{J}_x \rho := L_x \rho L_x^\dagger\,.
\end{equation}
Following the jump measurement $x$, the state $\rho$ becomes
\begin{equation}
    \rho \overset{x}{\mapsto} \frac{\mathcal{J}_x \rho}{\tr\{ \mathcal{J}_x \rho \}}\,,
\end{equation}
where the trace corrects for normalization. One can also describe the \keyword{no-jump superoperator},
\begin{equation}
    \mathcal{L}_0 := \mathcal{L} - \sum_{x\in\mathbb{M}} \mathcal{J}_x\,,
\end{equation}
to describe evolution in the absence of observed quantum jumps. With the initial state $\rho$, the probability of measuring a given trajectory is given by~\cite{landi_patterns_2023}
\begin{equation}\label{eq:prob_traj}
    \probof{\tau_1,x_1,\ldots,\tau_{\numMeas},x_{\numMeas}} = \tr\left\{ \mathcal{J}_{x_{\numMeas}}e^{\mathcal{L}_0 \tau_{\numMeas}} \cdots \mathcal{J}_{x_1}e^{\mathcal{L}_0 \tau_1} \rho \right\}\,.
\end{equation}

If we only care about the sequence of jumps and not the timing between them, we can integrate the probability over each $\tau_i\in[0,\infty)$. After integrating, the no-jump part becomes  $\int_0^\infty e^{\mathcal{L}_0 \tau}d\tau = -\mathcal{L}_0^{-1}$. The term from the upper bound of the integral vanishes as long as there are no \keyword{dark states}; that is, a jump will always be observed at some time in the future for any initial state. 
The probability of observing a sequence of jumps with arbitrary time intervals between them is then
\begin{equation}\label{eq:prob_seq}
    \probof{x_1,\ldots,x_{\numMeas}} = \tr\{\mathcal{M}_{x_{\numMeas}} \cdots \mathcal{M}_{x_1}\rho\}\,,
\end{equation}
for the superoperator,
\begin{equation}
    \mathcal{M}_x = -\mathcal{J}_x\mathcal{L}_0^{-1}\,.
\end{equation}
These superoperators define the quantum instrument for observing jumps without time tags.

The ``steady state'' of interest for the quantum jump process [Eq.~\eqref{eq:master_eq}] is the so-called \keyword{jump steady state} (JSS)\footnote{Note that this should not be confused with the \keyword{steady state} resulting describing the state asymptotic state of a system following a Lindblad master equation; i.e., $\frac{d}{dt}\rho_{\textrm{ss}} = \mathcal{L}\rho_{\textrm{ss}}=0$. 
The two are related as~\cite{landi_patterns_2023}:
\begin{equation*}
    \pi = \frac{\mathcal{J}\rho_{\textrm{ss}}}{K}\quad\textrm{for}\quad K=\tr\left\{ \mathcal{J}\rho_{\textrm{ss}} \right\}\,.
\end{equation*}
Though this steady state $\rho_{\textrm{ss}}$ is common throughout the broader literature, only $\pi$ is used in this work.
} satisfying Eq.~\eqref{eq:instrument_steady}: $\mathcal{M}\pi=\pi$.

\subsection{Relative entropy}
In this section, we will develop a method for quantifying the effects of measurement backaction and correlations. A crucial tool for this is the \keyword{Kullback-Leibler (KL) divergence}, or \keyword{relative entropy}~\cite{kullback_information_1951, yuan_hypothesis_2019},
\begin{equation}\label{eq:kldiv}
    \kldiv{\mathcal{P}}{\mathcal{Q}} = \int \mathcal{P}(z) \log\left(\frac{\mathcal{P}(z)}{\mathcal{Q}(z)}\right) dz\,,
\end{equation}
where $\mathcal{P}$ and $\mathcal{Q}$ are probability distributions with the same support and the integral is over the entire support. Analogous to the idea of \keyword{mutual information}, we will consider the relative entropy between the case with and without correlations between measurements. 
In particular, we consider how the correlations affect the statistics of the empirical distribution.

For $d$-dimensional normal distributions $\mathcal{N}_1$ and $\mathcal{N}_2$ with covariance $\Sigma_1$ and $\Sigma_2$, respectively, separated by $\boldsymbol{\Delta\mu}$, the KL-divergence is
\begin{align}
    \kldiv{\mathcal{N}_1}{\mathcal{N}_2} &= \frac{1}{2}\left[ \tr\left( \Sigma_2^{-1} \Sigma_1 \right) + \boldsymbol{\Delta\mu}^\trans \Sigma_2^{-1} \boldsymbol{\Delta\mu} \right.\nonumber \\
    &\qquad\left. + \log\left( \frac{\det{\Sigma_2}}{\det{\Sigma_1}} \right) - d \right]\,. \label{eq:kldiv_gauss}
\end{align}
For convenience of convention, the KL-divergence between two normal distributions with variances $\Sigma_1$ and $\Sigma_2$ centered at the same point ($\boldsymbol{\Delta\mu}=0$) will be denoted as $\kldiv{\Sigma_1}{\Sigma_2}$. Note that this case is scale-invariant, $\kldiv{a\Sigma_1}{a\Sigma_2}=\kldiv{\Sigma_1}{\Sigma_2}$.

\subsection{Fisher information}
The \keyword{Fisher information} is a useful metrological quantity as it provides a lower-bound on the uncertainty of a parameter achievable for a given set of measurements. 
The main calculation needed for the KL-divergence described in the previous section is likewise the main calculation needed for the Fisher information; namely the $\Psi$-matrix [e.g., Eq.~\eqref{eq:PsiMat}]. Thus, the work in this paper to calculate information theoretic quantities directly translates to calculating this metrological quantity.

For the Fisher information in general, denote a set of data $\boldsymbol{X}$ sampled from a set of possible measurements $\Omega$ according to the distribution $\probof{\boldsymbol{X}}$ parameterized by the variable $\theta$ (for simplicity, we will assume $\theta\in\mathbb{R}$ throughout); note that the Fisher information can be described independent of the data structure, but we specifically consider $\boldsymbol{X}$ describing a sequence of measurements in this work. The Fisher information for estimating $\theta$ from the data set $\boldsymbol{X}$ is
\begin{equation}
    F_\theta = \sum_{\boldsymbol{X}\in\Omega} \frac{\left[ \partial_\theta \probof{\boldsymbol{X}} \right]^2}{\probof{\boldsymbol{X}}}\,.
\end{equation}
This quantity is entirely independent of the realization of a measurement outcome. 

The utility of Fisher information can be seen in the \keyword{Cram\'er-Rao bound} on the variance~\cite{rao_information_1992, kay_cramer-rao_2013},
\begin{equation}
    \textrm{Var}\left[ Q\left( \boldsymbol{X} \right) \right] \geq \frac{\left\{ \partial_\theta \expect{Q\left( \boldsymbol{X} \right)} \right\}^2}{F_{\theta}}\,,
\end{equation}
where $Q$ is an arbitrary function of the data. In particular, one is often interested in the case where $Q\left( \boldsymbol{X} \right) = \hat{\theta}\left( \boldsymbol{X} \right)$ is an estimator for parameter $\theta$ using data $\boldsymbol{X}$, $\hat{\theta}\left( \boldsymbol{X} \right)\approx \theta$, and $\expect{Q\left( \boldsymbol{X} \right)} = \sum_{\boldsymbol{X}} \probof{\boldsymbol{X}} Q\left( \boldsymbol{X} \right)$ is the expected value over possible measurements. 
Note that $\partial_\theta \expect{\hat{\theta}\left( \boldsymbol{X} \right)}=1$ for an \keyword{unbiased estimator}. 

The Fisher information for a Gaussian distribution is well-understood:
\begin{align}\label{eq:FI_gauss}
    F_\theta^{\textrm{(Gauss)}} &= \left( \partial_\theta \boldsymbol{\mu} \right)^\trans \Sigma^{-1} \left( \partial_\theta \boldsymbol{\mu} \right) \nonumber \\
    &\qquad + \frac{1}{2} \tr\left\{ \Sigma^{-1} \left(\partial_\theta\Sigma\right) \Sigma^{-1} \left(\partial_\theta\Sigma\right) \right\} \,,
\end{align}
where the mean $\boldsymbol{\mu}$ and covariance matrix $\Sigma$ characterize the Gaussian. 
For the distribution of empirical distributions, $\boldsymbol{\mu}=\boldsymbol{p}:=\{p_x\}$ and $\Sigma$ is given by Eq.~\eqref{eq:empdist_covar}. Observe that the first term is $\mathcal{O}(\numMeas)$ while the second term is $\mathcal{O}(1)$, so one can ignore the second term in the large-$\numMeas$ limit. Thus, for large $\numMeas$, the Fisher information of the empirical distribution is
\begin{equation}\label{eq:FI_emp_discrete}
    F_\theta^{\textrm{(emp)}} = \numMeas \left( \partial_\theta \boldsymbol{p} \right)^\trans \left( \mathbb{P} + \Psi\mathbb{P} + \mathbb{P}\Psi^\trans \right)^{-1} \left( \partial_\theta \boldsymbol{p} \right)\,,
\end{equation}
where $\mathbb{P}=\textrm{diag}\,\boldsymbol{p}$. Note that Eq.~\eqref{eq:FI_emp_discrete} is a bit different than plugging Eq.~\eqref{eq:empdist_covar} into Eq.~\eqref{eq:FI_gauss} due to vanishing terms; an explanation of this is found in the Appendices of Ref.~\cite{smiga_stochastic_2023}.

In the case of \keyword{independent and identically distributed} (iid) measurements, $\probof{X_1,\ldots,X_{\numMeas}}=\prod_{i=1}^{\numMeas} \probof{X_i}$, one would expect the empirical distribution to give just as much information as the full sequence. That is, there is no additional information in the order of the independent measurements. In fact, one observes $\Psi=0$ in this case, so
\[
F_\theta^{\textrm{(emp)}} = \numMeas \sum_{x\in\mathbb{M}} \frac{\left( \partial_\theta p_x \right)^2}{p_x}\,,
\]
as expected.

\section{Influence of correlations on the empirical distribution for a quantum instrument}\label{sec:empCorrInfo_quant}

Consider an empirical distribution for a quantum jump process; i.e., a histogram of the observed jumps. We are interested in quantifying the role correlations play in the empirical distribution. As shown in Eq.~\eqref{eq:empdist_covar}, this is encoded in the matrix $\Psi$. The only remaining step is to apply the quantum instruments statistics detailed in Sec.~\ref{sec:quantum_instruments}. Using this information, one can easily estimate the relative entropy, Fisher information, or similar statistical quantity; assuming that the distribution of empirical distributions can be approximated as a Gaussian.

A detailed calculation of the effects of correlation for a quantum jump process is given in Appendix~\ref{app:PsiMat}. The main result of this is Eq.~\eqref{eq:app_PsiMat_quantJmps}:
\begin{equation}\label{eq:PsiMat_quantJmps}
    \Psi_{yx} \overset{\numMeas\gg1}{\approx} \frac{1}{p_x} \bbraopkket{1}{\mathcal{M}_y (\ident-\mathcal{M})^\draz \mathcal{M}_x}{\pi}\,,
\end{equation}
where $\mathcal{A}^\draz$ denotes the Drazin inverse of $\mathcal{A}$. 
Here, we use the \keyword{vectorized} notation where density matrices $\pi$ are treated as vectors $\kket{\pi}$ and the inner product is given by $\bbrakket{\sigma}{\rho}=\tr{\sigma^\dagger\rho}$.
One can extend this to the general ED$\seqLen$, as shown in Appendix~\ref{app:seq_measure}, yielding Eq.~\eqref{eq:app_PsiMat_quantJmps_seq}:
\begin{align}\label{eq:PsiMat_quantJmps_seq}
    \Psi_{\boldsymbol{y}\boldsymbol{x}} &\overset{\numMeas\gg\seqLen}{\approx} \frac{1}{p_{\boldsymbol{x}}} \bbra{1}{\mathcal{M}_{\boldsymbol{y}}\left(1-\mathcal{M}\right)^{+}\mathcal{M}_{\boldsymbol{x}}}\kket{\pi} \nonumber \\
    & \qquad +  \sum_{\ell=1}^{\seqLen-1} \left[ p_{\boldsymbol{y}\gets \boldsymbol{x}}(\ell)-p_{\boldsymbol{y}} \right]\,.
\end{align}
This extension to finite sequences encodes the role of correlations in strings of data instead of simply individual measurements, though it requires some additional terms due to the effect of the subsequences overlapping; i.e., the second line of Eq.~\eqref{eq:PsiMat_quantJmps_seq}.
For example, if $\seqLen=2$, $\boldsymbol{x}=x_1x_2$, and $\boldsymbol{y}=y_1y_2$
\[
    p_{\boldsymbol{y}\gets \boldsymbol{x}}(1)-p_{\boldsymbol{y}} = \bbra{1} \mathcal{M}_{y_2}\mathcal{M}_{y_1} \left( \frac{\delta_{x_2y_1}\mathcal{M}_{x_1}}{p_{\boldsymbol{x}}} - 1 \right) \kket{\pi}\,.
\]
Similar for $\seqLen=3$,
\begin{align*}
\sum_{\ell=1}^{2} &\left[ p_{\boldsymbol{y}\gets \boldsymbol{x}}(\ell)-p_{\boldsymbol{y}} \right] = \\
&\bbra{1}{\mathcal{M}_{y_3} \mathcal{M}_{y_2}\mathcal{M}_{y_1}\left[ \frac{\delta_{y_2x_3}\delta_{y_1x_2} \mathcal{M}_{x_1}}{p_{\boldsymbol{x}}} - 1\right]}\kket{\pi} \\
&+ \bbra{1}{\mathcal{M}_{y_3}\mathcal{M}_{y_2} \mathcal{M}_{y_1}\left[\frac{\delta_{y_1x_3} \mathcal{M}_{x_2}\mathcal{M}_{x_1}}{p_{\boldsymbol{x}}} - 1\right] }\kket{\pi}\,.
\end{align*}
Though perhaps not as ``clean'' as the $\seqLen=1$ case, there are finitely many such additional terms, even in the large-$\numMeas$ limit.

\section{Empirical distribution mutual information}\label{sec:relEnt_empDist}

\subsection{Distribution of measurements}
The empirical distribution provides a convenient way to understand the effect of measurement backaction on the correlations. By itself, the empirical distribution differs from the full data by omitting information concerning the order of observations. Thus, it would seem to also omit the effects of correlations. However, as seen in Eq.~\eqref{eq:empdist_covar}, this information is efficiently encoded in the covariance of the empirical distribution. 
Separating the covariance into the components corresponding to uncorrelated and correlated components,
\begin{subequations}\label{eq:covar_decomp}
    \begin{align}
        \Sigma &= \frac{1}{N} \Sigma_0 \\
        \Sigma_0 &= \Sigma_P + \Sigma_\Psi \\
        \Sigma_P&=\mathbb{P}-\boldsymbol{p}\boldsymbol{p}^\trans \\
        \Sigma_\Psi &=\Psi\mathbb{P} + \mathbb{P}\Psi^\trans\,.
    \end{align}
\end{subequations}

Note that the covariance matrix $\Sigma$, as written, is not positive-definite. This is reflective of ``overfitting'' the empirical distribution after normalizing for the number of measurements. The nullspace of these three matrices consists of a vector with identical entries --- assuming an ergodic process, this will be the entire nullspace for $\Sigma$ and $\Sigma_P$. Likewise, variation in the empirical distribution in this direction reflects a break in the normalization. 
It will be useful to define a mathematical tool, the \keyword{pseudo-determinant},
\begin{equation}
    \pdet{\Sigma} := \prod_{\lambda_i\neq0} \lambda_i\,,
\end{equation}
which describes the determinant in the subspace orthogonal to the nullspace.
If we define the matrix $N$ to have columns that form an orthonormal basis of the nullspace of $\Sigma$, then $\pdet\Sigma = \det\left( \Sigma+N N^\trans \right)$. The addition of $N N^\trans$ serves the role of transforming the nullspace of $\Sigma$ into a 1-eigenspace without affecting the other eigenvalues.
When using Eq.~\eqref{eq:kldiv_gauss} to calculate the KL-divergence, this will be the relevant determinant to use. This is a result of only integrating over the domain of support in Eq.~\eqref{eq:kldiv}.
Similarly, the inverse should be replaced with the Drazin inverse, which has the effect of replacing the kernel with a 1-eigenspace, taking the inverse, then switching back the kernel; c.f., Appendix~L of Ref.~\cite{landi_current_2024} for more information regarding the Drazin inverse.

The correlations between measurements are encoded in $\Psi$, so taking $\Psi\to0$ (hence $\Sigma_\Psi\to 0$) is equivalent to removing the influence of correlations. This is equivalent to taking measurements from independent systems to build the empirical distribution. 
We thus consider the following quantity to describe the information contained in the correlations, 
\begin{align*}
\mathcal{I} &= \kldiv{\Sigma_0}{\Sigma_P} \\
&= \frac{1}{2}\left[ \tr\left( \Sigma_P^\draz \Sigma_0 \right) + \log\left( \frac{\pdet{\Sigma_P}}{\pdet{\Sigma_0}} \right) - d \right] \\
&= \frac{1}{2}\left[ \tr\left( \ident + \Sigma_P^\draz \Sigma_\Psi \right) + \log\left( \frac{\pdet{\Sigma_P}}{\pdet{\Sigma_0}} \right) - d \right]\,.
\end{align*}
As shown in Appendix~\ref{app:ident_probdet}, $\pdet{\Sigma_P}=d\prod_{i=1}^d p_i$.
Recalling also that $\tr\left(\ident\right)=d$, this results in 
\begin{align}\label{eq:corrInfo}
    \mathcal{I} &= \frac{1}{2}\left[ \log\frac{\pdet{\Sigma_P}}{\pdet{\Sigma_0}} + \tr\left( \Sigma_P^\draz \Sigma_\Psi \right) \right]\\
    &= \frac{1}{2}\left[ \log\frac{d}{\pdet{\Sigma_0}} + \sum_{i=1}^{d} \log\left( p_i \right) + \tr\left( \Sigma_P^\draz \Sigma_\Psi \right) \right]\,. \nonumber
\end{align}
Some properties immediately follow; for example, $\mathcal{I}=0$ for iid measurements.

\subsection{Distribution of finite sequences}\label{sec:info_finiteSeq}
We are now interested in extending the work outlined in the previous section to the empirical distribution of finite sequences [Eq.~\eqref{eq:empdist_seq_general}]. 

Na\"ively, one might consider omitting correlations by simply imagining an experiment which resets after every $\seqLen$ measurements with each experiment corresponding to a single observation. This would be a natural iid model that closely reflects the full sequence.
However, this will be immediately distinguishable from the case in which the experiment does not reset, because the latter case has a much smaller support. 
The constraints are made explicit in Appendix~\ref{app:seq_constraints}, though the basic idea is that a system must ``enter'' a state the same number of times it ``leaves.'' 
The same strict constraint does not apply if each of the $\seqLen$-sequence are from different processes.

An alternative approach is to extend the model to which the process is compared beyond an iid approximations. A natural choice for this is the order-$\markOrd$ Markov process. When $\markOrd=0$, one recovers an iid process, while higher orders enables correlations between finitely separated measurements. 
The goal in this section, then, will be to extend the main result in the previous section (namely Eq.~\eqref{eq:corrInfo}) to compare the full system model to one marginalized to a process of arbitrary order,
\[
\kldiv{\Sigma_0}{\Sigma_P} \to \kldiv{\Sigma_0}{\numMeas \Sigma_{(\markOrd,\seqLen)}}\,,
\]
where $\Sigma_{(\markOrd,\seqLen)}$ is the covariance of the distribution of empirical distributions consisting of $\seqLen$-sequences marginalized to a $\markOrd$-order Markov process.

Observe that the empirical distribution itself encodes similar information as the Markov matrix (or tensor for $\markOrd>1$). 
One finds the average empirical distribution is $\expect{q_{\boldsymbol{x}}}=\probof{\boldsymbol{x}}$, and the Markov tensor is $Q_{x_1\ldots x_\markOrd}^{x_{\markOrd+1}} = \frac{\probof{x_1,\cdots,x_{\markOrd+1}}}{\probof{x_1,\cdots,x_\markOrd}}$. When $\markOrd+1\geq \seqLen$, the average empirical distribution can be derived from the steady state of the Markov process, $p_{x_1\ldots x_\markOrd} = \probof{x_1,\ldots, x_\markOrd}$.
One can take a general theory of a stationary process, marginalize the statistics to an order-$\markOrd$ Markov process --- in which long-term correlations are suppressed --- and compare this to the original to gain quantitative insight into the role of correlations. Here, we consider the KL-divergence in the distribution of $\seqLen$-empirical distribution to provide such a quantity.

\begin{figure}[ht]
    \centering
    \includegraphics[width=.9\linewidth]{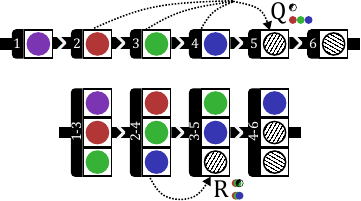}
    \caption{Visualization of $\markOrd=3$-order Markov process and how it can be reduced to a first-order process. Above is the process as observed and below is an order-1 process obtained by combining sets of three measurements into single elements in the sequence.}
    \label{fig:markovOrder}
\end{figure}

If one chooses $\markOrd+1\geq\seqLen$, then the finite-sequence empirical distribution does not ``overfit'' the Markov model. As a result, the average empirical distribution for the original model and the Markov model will be the same, so the two models are only distinguishable by the covariance of the empirical distributions. Because both covariance matrices scale inversely with the $\numMeas$ measurements, the distinguishability of the models does not depend on $\numMeas$, as long as they are centered around the same mean.

The Markov order $\markOrd=1$ case was explored in Ref.~\cite{smiga_stochastic_2023}, but this can be easily extended to the general case. Using Eq.~\eqref{eq:empdist_covar} to calculate the covariance, $\boldsymbol{p}$ is the steady-state. When $Q$ is the Markov matrix, the covariance is encoded in
\[
\Psi = Q(1-Q)^\draz\,.
\]
Extending to $\seqLen>1$ follows by indexing the vectors and matrices by sequences of measurements. For example, $p_x\to p_{x_1\ldots x_{\seqLen}} = \probof{x_1,\ldots, x_{\seqLen}}$ and $Q_{x}^{y}\to Q_{x_1\ldots x_{\seqLen}}^{y_2\ldots y_{\seqLen+1}} = \delta_{x_2y_2}\cdots\delta_{x_{\seqLen}y_{\seqLen}}Q_{x_{\seqLen}}^{y_{\seqLen+1}}$.
Extending to higher Markov orders follows a similar prescription. Define a matrix $R$ indexed by $\markOrd$-sequences using the Markov tensor,
\begin{equation}\label{eq:reduce_markov}
    R_{x_1\ldots x_\markOrd}^{y_2\ldots y_{\markOrd+1}} = \delta_{x_2y_2}\cdots \delta_{x_\markOrd y_\markOrd}Q_{x_1\ldots x_\markOrd}^{y_{\markOrd+1}}\,.
\end{equation}
This matrix acts on the Markov model steady state $p_{x_1\ldots x_\markOrd} = \probof{x_1,\ldots, x_\markOrd}$ as $R\boldsymbol{p}=\boldsymbol{p}$. 
As illustrated in Fig.~\ref{fig:markovOrder}, this new matrix effectively reduces the $\markOrd$-order process to a first-order process.
It is straightforward to show that the correlations are now encoded in the matrix $\Psi = R(1-R)^\draz$. As before, one can extend this to empirical distributions of finite sequences.
Together, the covariance for an empirical distribution of $\seqLen$-sequences from an order-$\markOrd$ Markov process is
\begin{align}
    \Sigma_{(\markOrd,\seqLen)} = \frac{1}{\numMeas} \left[ \mathbb{P}-\boldsymbol{p}\boldsymbol{p}^\trans \right. &+ R(1-R)^\draz\mathbb{P} \nonumber \\
    &\left. + \mathbb{P}(1-R^\trans)^\draz R^\trans\right]\,.
\end{align}

For an arbitrary stationary process that generates an $\seqLen$-empirical distribution $\mathcal{P}_\seqLen$, one can marginalize this process into an $\markOrd$-order Markov process $\mathcal{Q}_{\seqLen}^{\markOrd}$. Then, the KL-divergence between these two processes,
\begin{align}\label{eq:corrInfo_general}
    \mathcal{I}_{\seqLen}^{\markOrd} &= \kldiv{\mathcal{P}_\seqLen}{\mathcal{Q}_{\seqLen}^{\markOrd}} \nonumber \\
    &= \kldiv{\Sigma}{\Sigma_{(\markOrd,\seqLen)}}
\end{align}
provides a measure of agreement. Here, $\Sigma$ is calculated from Eq.~\eqref{eq:empdist_covar} using $\Psi$ from Eq.~\eqref{eq:PsiMat_quantJmps_seq}. If $\mathcal{P}$ is a Markov process, $\mathcal{I}_{\seqLen}^{\markOrd}$ will vanish as long as $\markOrd$ is at least the Markov order of $\mathcal{P}$. Also, if $\seqLen>\markOrd+1$, then $\mathcal{P}_\seqLen$ and $\mathcal{Q}_{\seqLen}^{\markOrd}$ will be centered at different means, so $\mathcal{I}_{\seqLen}^{\markOrd}$ diverges in the large-$\numMeas$ limit. 

\medskip
\centerline{
    \rule[2.5pt]{0.1\linewidth}{0.5pt}%
    \rule[2.25pt]{0.1\linewidth}{1pt}%
    \rule[2pt]{0.2\linewidth}{1.5pt}%
    \rule[2.25pt]{0.1\linewidth}{1pt}%
    \rule[2.5pt]{0.1\linewidth}{0.5pt}%
}
\bigskip

At this point, we have demonstrated several tools for quantifying the information for a general quantum process encoded in the empirical distribution; a simple counting statistic. Equations~\eqref{eq:PsiMat_quantJmps} and~\eqref{eq:PsiMat_quantJmps_seq} encode the role correlations play in the empirical distribution. Using this quantity to calculate the covariance, Eq.~\eqref{eq:empdist_covar}, one can explore additional information theoretic (e.g., the relative entropy, Eqs.~\eqref{eq:corrInfo} and~\eqref{eq:corrInfo_general}) and meteorological (e.g., Fisher information, Eq.~\eqref{eq:FI_emp_discrete}) quantities.
Additional considerations were needed to extend this work to empirical distributions of sequences of measurements, though one can, e.g., calculate the relative entropy in such data as compared to the model marginalized to a Markov processes of sufficient order, Eq.~\eqref{eq:corrInfo}.
In the following section, we apply these methods to illustrative examples.

\section{Examples}\label{sec:examples}

\begin{figure*}[ht]
    \centering
    \includegraphics[width=0.8\textwidth]{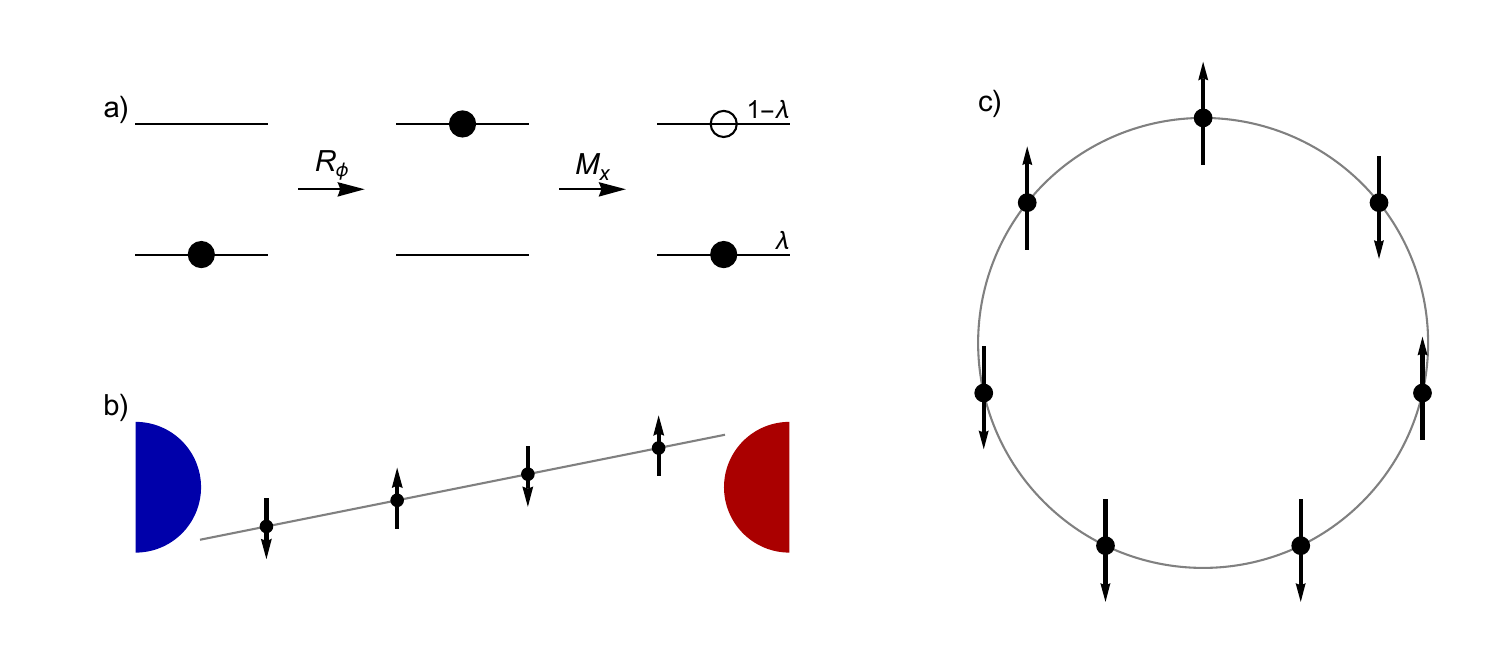}
    \caption{Illustrations of examples explored in this paper. (a) An amplitude-damped described in Sec.~\ref{sec:ampDamp}. (b) A boundary-driven spin chain described in Sec.~\ref{sec:xx_chain}. (c) A periodic spin chain described in Sec.~\ref{sec:periodic_chain}.}
    \label{fig:examples}
\end{figure*}

Various examples using quantum systems are explored in this section. Figure~\ref{fig:examples} summarizes the systems considered.

\subsection{Amplitude damped qubit}\label{sec:ampDamp}
As a simple toy model, consider a qubit undergoing repeated amplitude-damping-type observations via the Kraus operators,
\begin{subequations}\label{eq:ampDampKraus}
    \begin{align}
        M_0 &= \left[ \begin{matrix}
            1 & 0 \\
            0 & \sqrt{1-\lambda}
        \end{matrix} \right] \\
        M_1 &= \left[ \begin{matrix}
            0 & \sqrt{\lambda} \\
            0 & 0
        \end{matrix} \right]\,,
    \end{align}
\end{subequations}
for $\lambda\in[0,1]$. 
Applying this as a channel, $\rho\mapsto M_0\rho M_0^\dagger+M_1\rho M_1^\dagger$, acts similar spontaneous emission when $\lambda=1$; the system is always left in the ground state after applying the channel. 
When $\lambda<1$, the system is not fully reset.

Note that the Kraus decomposition is not unique. In general, one can use the operators,
\begin{align*}
 N_0 &= \left[ \begin{matrix}
            e^{i \alpha } \cos (\theta ) & e^{i \beta } \sqrt{\lambda } \sin (\theta ) \\
            0 & e^{i \alpha } \sqrt{1-\lambda } \cos (\theta ) \\
        \end{matrix} \right] \\
N_1 &= \left[ \begin{matrix}
            -e^{-i \beta } \sin (\theta ) & e^{-i \alpha } \sqrt{\lambda } \cos (\theta ) \\
            0 & -e^{-i \beta } \sqrt{1-\lambda } \sin (\theta )
        \end{matrix} \right]\,,
\end{align*}
for $\beta,\theta\in\mathbb{R}$. In this section, we choose to use Eq.~\eqref{eq:ampDampKraus} to provide a convenient toy model.

For any non-zero $\lambda$, the state will tend towards the ground state over time. To make the dynamics of the system a bit more interesting, an $X$-rotation can be inserted between observations,
\begin{equation}
    R_X(\phi)=e^{-i\frac{\phi}{2}\sigma_x}=\left[ \begin{matrix}
        \cos\frac{\phi}{2} & -i\sin \frac{\phi}{2} \\
        -i\sin \frac{\phi}{2} & \cos\frac{\phi}{2}
    \end{matrix}\right]\,.
\end{equation}
This operation prevents the sequence of observations from converging to ``$M_0$'' by allowing the state to be systematically ``scrambled'' between observations. 

The dynamics of the system are described by the measurement superoperators, 
\begin{equation}\label{eq:ampDamp_instruments}
    \mathcal{M}_j\rho = M_jR_X(\phi)\rho R_X(\phi)^\dagger M_j^\dagger\quad\textrm{for}\quad j\in\{0,1\}\,.
\end{equation}
This model is parameterized by two numbers: $\lambda\in[0,1]$ and $\phi\in[0,2\pi)$. 

\begin{figure}[ht]
    \centering
    \includegraphics[width=\columnwidth]{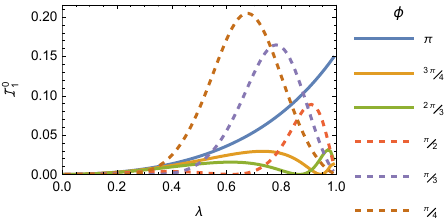}
    \caption{Information contained in correlations between measurements for the amplitude-damping qubit example for different damping $\lambda$ and intermittent rotations $\phi$.}
    \label{fig:ampDamp_corrInfo}
\end{figure}

Consider first the special case with $\phi=\pi$, so $R_X(\pi)=-i\sigma_x$ which has the effect of a total bitflip, up to a global phase. 
In this case, if one starts in the pure excited state $\rho_e=\left[\begin{matrix}0&0\\0&1\end{matrix}\right]$, one will guarantee that $x=0$ is observed and the system ends in the ground state $\rho_g=\left[\begin{matrix}1&0\\0&0\end{matrix}\right]$. However, when starting in the ground state, there is a $\lambda$ probability of measuring $x=1$ and remaining in the ground state, while a $1-\lambda$ chance of measuring $x=0$ and flipping to the excited state. Since the off-diagonal terms are suppressed by a factor of $\sqrt{1-\lambda}$ when $x=0$ is measured --- they also do not appear in the steady state --- they do not play a significant role in the large-$\numMeas$ limit. Thus, together, this special case can be described by an ``even machine'' wherein an even number of ``0''s are measured between ``1''s.

For the bitflip case, there are relatively simple analytic solutions to many of the quantities of interest in this work. First, the steady state is given by
\begin{equation}
    \pi_{\textrm{SS}} = \left[\begin{matrix}\frac{1}{2-\lambda}&0\\0&\frac{1-\lambda}{2-\lambda}\end{matrix}\right]\,.
\end{equation}
The average distribution of outcomes is given by (in $\{0,1\}$-ordered basis)
\begin{equation}
    \boldsymbol{p} = \left[ \begin{matrix}
        \frac{2(1-\lambda)}{2-\lambda} \\ \frac{\lambda}{2-\lambda}
    \end{matrix} \right]\,.
\end{equation}
Evaluating Eq.~\eqref{eq:PsiMat_quantJmps}, one finds
\begin{equation}
    \Psi = \frac{\lambda}{(2-\lambda)^2} \left[ \begin{matrix}
        \frac{\lambda}{2} & -(1-\lambda) \\
        -\frac{\lambda}{2} & (1-\lambda) 
    \end{matrix} \right]\,.
\end{equation}
Then, the covariance (decomposed from Eq.~\eqref{eq:covar_decomp}) are given by
\begin{equation}
    \Sigma_0 = \frac{4(1-\lambda)\lambda}{(2-\lambda)^3} \left[\begin{matrix}1&-1\\-1&1\end{matrix}\right]\quad\textrm{and}\quad\Sigma_P=\frac{2-\lambda}{2} \Sigma_0\,.
\end{equation}
One can quantify the information in the correlations between measurements, Eq.~\eqref{eq:corrInfo}, as 
\begin{equation}
    \mathcal{I} = \frac{1}{2} \left[ \frac{\lambda}{2-\lambda} + \log \left(\frac{2-\lambda}{2}\right) \right]\,.
\end{equation}
This case, along with other intermittent rotations is shown in Fig.~\ref{fig:ampDamp_corrInfo}. 

The system in the case of bitflips is equivalent to an ``even machine.'' In particular, each pair of ``0''s occurs with probability $(1-\lambda)$. As $\lambda$ approaches 1, the probability of measuring ``0'' decreases, however, when a single ``0'' is measured, one is guaranteed to observe a second ``0.'' If we were to marginalize to an approximate iid model, this doubling would not be observed. 
In contrast, for small $\lambda$, the lengths of ``0'' strings tend to be very large, so the fact that ``0''s always appear in pairs does not have such a dramatic influence on the statistics. Thus, one sees that $\mathcal{I}$ increases with $\lambda$ in Fig.~\ref{fig:ampDamp_corrInfo}.
For the case of general rotations, additional features can be observed due to different ``resonances.''

\begin{figure}[ht]
    \centering
    \includegraphics[width=\columnwidth]{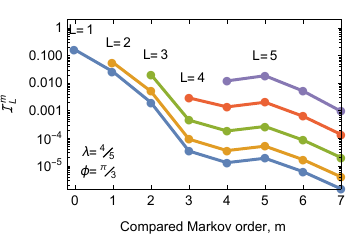}
    \caption{The KL-divergence of $\seqLen$-sequences for the amplitude-damped qubit, described by Eq.~\eqref{eq:ampDamp_instruments} as compared to the process marginalized to a Markov order-$\markOrd$ process. The damping is $\lambda=\frac{4}{5}$ and a $\phi=\frac{\pi}{3}$ rotation is applied between measurements.}
    \label{fig:ampDamp_empInfo_LM}
\end{figure}

Likewise, the information in correlations for a given set of parameters when comparing to general Markov of finite-sequences [Eq.~\eqref{eq:corrInfo_general}] is summarized in Fig.~\ref{fig:ampDamp_empInfo_LM}. In general, one finds that shorter sequences and higher Markov order are more difficult to distinguish from the full processes (i.e., $\mathcal{I}_{\seqLen}^{\markOrd}$ is smaller). This follows from the idea that higher-order Markov processes are more capable of encoding complicated correlations and longer sequences can capture additional subtleties. The non-vanishing of $\mathcal{I}$ supports the notion that this is a non-Markovian process. That is, if the data could be modeled by a Markov process, then matching the data to a model with sufficient high Markov order should lead to a vanishing KL-divergence when comparing the model.

\subsection{Boundary-driven $XX$-spin chain}\label{sec:xx_chain}
In this section, we consider the \keyword{boundary-driven $XX$-spin chain} detailed, e.g., in Refs.~\cite{landi_flux_2014, schuab_energy_2016, silva_nontrivial_2023, landi_patterns_2023}. In this model, a length-$\chainLen$ spin chain with hopping terms is connected to the environment on the boundaries. The environment is able to cause the spins on ends of the chain to flip. 
The Hamiltonian for the spin chain is given by
\begin{equation}
    H = J\sum_{i=1}^{\chainLen-1} \left( \sigma_i^{+} \sigma_{i+1}^{-} + \sigma_i^{-} \sigma_{i+1}^{+} \right) + \sum_{i=1}^{\chainLen} h_i \sigma_i^z\,,
\end{equation}
where $\sigma_i^\pm$ are Pauli operators for the $i^\textrm{th}$ site. Here, we will consider the case where $\{h_i\}$ varies linearly from $-\frac{h}{2}$ to $+\frac{h}{2}$: $h_j = h \frac{j-1}{\chainLen-1}-\frac{h}{2}$.
The Lindblad master equation for this system is
\begin{align}
    \frac{d\rho}{dt} = \mathcal{L}\rho = -i\commute{H}{\rho} &+ \gamma_{\textrm{L}} f_{\textrm{L}} \mathcal{D}\left[\sigma_1^{+}\right]\rho \nonumber \\
    &+ \gamma_{\textrm{L}} (1-f_{\textrm{L}}) \mathcal{D}\left[\sigma_1^{-}\right]\rho \nonumber \\
    &+ \gamma_{\textrm{R}} f_{\textrm{R}} \mathcal{D}\left[\sigma_{\chainLen}^{+}\right]\rho  \nonumber \\
    &+ \gamma_{\textrm{R}} (1-f_{\textrm{R}}) \mathcal{D}\left[\sigma_{\chainLen}^{-}\right]\rho\,.
\end{align}
The four dissipators correspond to the jump superoperators $\mathcal{J}_{\textrm{L}+}$, $\mathcal{J}_{\textrm{L}-}$, $\mathcal{J}_{\textrm{R}+}$, and $\mathcal{J}_{\textrm{R}-}$, respectively.

\begin{figure}[ht]
    \centering
    \includegraphics[width=\columnwidth]{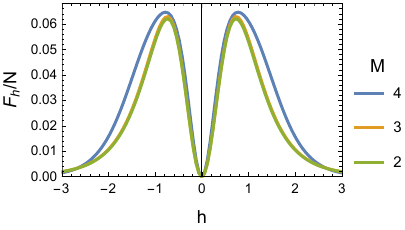}
    \caption{The rate of Fisher information per observation in the empirical distribution for boundary-driven, length-$\chainLen$, $XX$-spin chain with respect to external magnetic field gradient (varying from $-\frac{h}{2}$ to $\frac{h}{2}$ linearly across the spin chain). The observable jumps are emissions: $\mathcal{J}_{\textrm{L}-}$ and $\mathcal{J}_{\textrm{R}-}$. Units are chosen such that $J=1$. Other terms are set to be constant: $f_{\textrm{L}}=\frac{3}{4}$ and $f_{\textrm{R}}=\frac{1}{4}$ reflecting a finite spin-up bath on the left and spin-down bath on the right, and $\gamma_{\textrm{L}}=\gamma_{\textrm{R}}=1$. }
    \label{fig:bndryChain_FIRate}
\end{figure}

By observing the jumps, $\mathcal{J}_{\textrm{L}-}$ and $\mathcal{J}_{\textrm{R}-}$, one can estimate the gradient of the magnetic field $h$ across the spin chain. The Fisher information for $h$ in the empirical distribution from a chain with different lengths is shown in Fig.~\ref{fig:bndryChain_FIRate}. For the respective parameters ($J=\gamma_{\textrm{L}}=\gamma_{\textrm{R}}=1$ and $f_{\textrm{L}}=1-f_{\textrm{R}}=\frac{3}{4}$), the Fisher information maximizes on $h\sim\mathcal{O}(1)$. 

\begin{figure}[ht]
    \centering
    \includegraphics[width=\columnwidth]{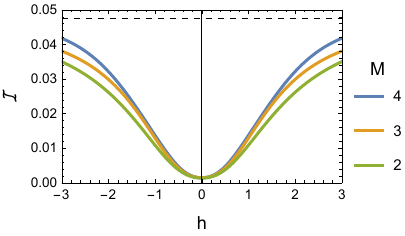}
    \caption{Correlational information encoded in the boundary-driven length-$\chainLen$, $XX$-spin chain with respect to different magnetic field gradients (varying linearly from $-\frac{h}{2}$ to $\frac{h}{2}$ across the spin chain). Only emissions $\mathcal{J}_{\textrm{L}-}$ and $\mathcal{J}_{\textrm{R}-}$ are observed. The dashed line is obtained for $\lvert h\rvert\to\infty$. Units are chosen such that $J=1$. Other terms are set to be constant: $f_{\textrm{L}}=\frac{3}{4}$ and $f_{\textrm{R}}=\frac{1}{4}$ reflecting a finite spin-up bath on the left and spin-down bath on the right, and $\gamma_{\textrm{L}}=\gamma_{\textrm{R}}=1$.}
    \label{fig:bndryChain_corrInfo}
\end{figure}

Likewise, one can consider the amount of information contained in the correlations via the empirical distribution. As shown in Fig.~\ref{fig:bndryChain_corrInfo}, increasing the gradient of the magnetic field results in more information contained within the correlations. For $f_{\textrm{L}}=1-f$, $f_{\textrm{R}}=f$, and $\gamma_{\textrm{L}}=\gamma_{\textrm{R}}=1$, the information is bounded asymptotically by
\[
\mathcal{I} < \frac{1}{2} \log\left( \frac{1}{(1-f)^2+f^2} \right) - f(1-f)\,.
\]
One approaches this informational asymptote for large gradients $\lvert h \rvert\to\infty$.

\subsection{Periodic spin chain with random measurements}\label{sec:periodic_chain}
Similar to the $XX$-spin chain explored in the previous section, here we consider a spin chain with periodic boundary conditions: identifying site $\seqLen+k$ with site $k$. In addition to hopping term, we include a term corresponding to spontaneous pair creation/annihilation between adjacent sites,
\begin{align}
    H = - \sum_{i=1}^{\chainLen}&\left[ J\left( \sigma_i^{+} \sigma_{i+1}^{-} + \sigma_i^{-} \sigma_{i+1}^{+} \right) \right. \nonumber \\
    &\left. + \kappa \left( \sigma_i^{+} \sigma_{i+1}^{+} + \sigma_i^{-} \sigma_{i+1}^{-} \right) \right]\,.
\end{align}
When $\kappa\neq 0$, the total number of spins is not conserved.
Similar to Ref.~\cite{coppola_growth_2022}, this chain is projectively measured at random sites. Here, these measurements are performed with period $\tau$, resulting in the $2\chainLen$ instruments,
\begin{equation}\label{eq:periodic_chain_inst}
    \mathcal{M}_j^{\pm} \rho = \frac{1}{\chainLen} \sigma_j^{\pm}\sigma_j^{\mp} e^{-i\tau H} \rho e^{+i\tau H} \sigma_j^{\pm}\sigma_j^{\mp}\,.
\end{equation}
One can consider the two instruments one obtains when the information pertaining to the site is lost,
\begin{equation}\label{eq:periodic_chain_inst_nosite}
    \mathcal{M}^\pm = \sum_{j=1}^\chainLen \mathcal{M}_j^{\pm}\,.
\end{equation}

\begin{figure}[ht]
    \centering
    \includegraphics[width=0.9\columnwidth]{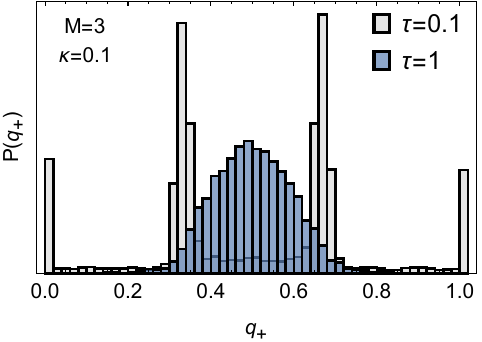}
    \caption{Distribution of empirical distributions for an $\chainLen=3$ periodic spin chain with random projective measurements (not knowing which site was measured) from $10^4$ simulations of $\numMeas=10^3$ measurement sequences. A small, $\kappa=0.1$ spin-non-preserving term is included and measurements are compared for rapid (gray) and gradual (blue) measurements. When the system is rapidly measured, the state of the system ``breaks'' towards a total spin number, while less frequent measurements allow the total number of spins to vary between measurements; the different number of peaks in the distributions is a result of the quantum Zeno effect. }
    \label{fig:prm_distdist}
\end{figure}

Interestingly, without including a spin-non-preserving term, this model would fail the assumption that the empirical distribution is sampled from a normal distribution, despite the process being stationary. If spins were preserved, then after sufficiently many measurements, one would gain a reasonable understanding of the total spin of the chain. This provides a type of ``strong memory'' that would persist over arbitrary time scales and be observed in the probabilities. Roughly, this splits the distribution of empirical distributions into a $(\chainLen+1)$-modal distribution, each corresponding to the system collapsing to a state with a given total spin (see Fig.~\ref{fig:prm_distdist}).

\begin{figure}[ht]
    \centering
    \includegraphics[width=\columnwidth]{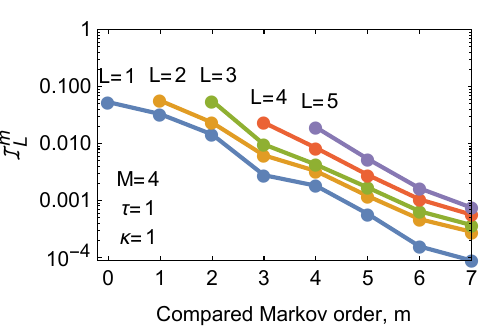}
    \caption{The KL-divergence of $\seqLen$-sequences for a length $M=4$ periodic spin chain with random projective measurements as compared to the process marginalized to a Markov order-$\markOrd$ process. The measurements do not distinguish between sites, as in Eq.~\eqref{eq:periodic_chain_inst_nosite}. 
    Units are such that $J=1$ and $\tau=\kappa=1$. }
    \label{fig:prm_empInfo_LM}
\end{figure}

Omitting site information, Fig.~\ref{fig:prm_empInfo_LM} summarizes the comparison of the KL-divergence between the model and corresponding Markov models for finite sequences. 
We observe a qualitatively similar trend as before: larger sequence lengths $\seqLen$ and smaller compared Markov order $\markOrd$ tend to result in a larger KL-divergence.

\section{Conclusion} \label{sec:conclusion}
In this work, we present methods for quantifying the role correlations play in making a sequence of measurements --- especially on quantum systems --- using the empirical distribution. Including correlations will lead to both information theoretic effects (seen in the relative entropy) and metrological effects (seen in the Fisher information). 
Understanding the role of correlations resulting from measurement backaction or other influences is useful to developing optimal and realistic measurement techniques; in addition to these effects being a natural result of measurements, expanding to a paradigm beyond ``prepare and measure,'' which restricts us to trivial correlations, may enable new, improved measurement schemes. 

There are a few immediate avenues of future work that could yield interesting results from the methods described in this manuscript. In addition to simply applying the formalism to various measurement schemes, our methods can potentially inform new methods that embrace correlations in measurements to improve calculations.
For example, recent work~\cite{prech_optimal_2024} explored the problem of time-keeping using a classical Markovian process, though this may be improved if considering more complicated processes.
In addition, it would be interesting to observe if effects like (measurement-induced) phase transitions could be observed in this formalism. By focusing on the empirical distribution, purely transient patterns are lost, allowing us to focus on persistent effects.

\bibliography{main_arxiv}

\appendix
\onecolumngrid

\section{$\Psi$ matrix for quantum jumps} \label{app:PsiMat}
In this appendix, we evaluate $\Psi$ for the case of quantum jumps. 
Throughout, we assume that the system is initially in the jump steady state, $\pi$, which is assumed to be unique. 
Further, we use the \keyword{vectorized notation} in which a density matrix $\rho$ is written as the vector $\kket{\rho}$, so superoperators are understood as linear operators acting on the space of operators. This forms an inner product space with the \keyword{Frobenius/Hilbert-Schmidt inner product},
\begin{equation}
    \bbrakket{\sigma}{\rho} = \tr\sigma^\dagger\rho\,.
\end{equation}
It follows that $\tr\rho=\bbrakket{1}{\rho}$, where $\bbra{1}$ is the covector associated with the identity operator.

\subsection{Single measurement empirical distribution} \label{app:PsiMat_single}
First, we consider the special case of empirical distributions with $\seqLen=1$ measurements. This is generalized to longer, finite-sequence case in the next section.

Recall that [Eq.~\eqref{eq:PsiMat}]
\begin{equation}\label{eq:app_PsiMat}
    \Psi_{yx} = \sum_{\ell=1}^{\numMeas-1} \left( 1-\frac{\ell}{\numMeas} \right)\left[ p_{y\gets  x}(\ell)-p_y \right]\,,
\end{equation}
for $p_y=\probof{y}$ and $p_{y\gets  x}(\ell)=\probgiven{X_{i+\ell}=y}{X_{i}=x}$. 
Using vectorized notation, 
\begin{align}
    p_y &= \bbraopkket{1}{\mathcal{M}_y}{\pi}\,, \\
    p_{y\gets  x}(\ell) &= \frac{1}{p_x} \bbraopkket{1}{\mathcal{M}_y\mathcal{M}^{\ell-1}\mathcal{M}_x}{\pi}\,,
\end{align}
for $\ell\geq 1$ and, for $\ell=1$, identifying $\mathcal{M}^{0}\to\ident$.

One can define an eigendecomposition of the superoperator $\mathcal{M}$ into projection superoperators:
\begin{equation}
    \mathcal{M} = \mathcal{P}_0 + \sum_{j\geq 1} \lambda_j \mathcal{P}_j\,,
\end{equation}
where $\mathcal{P}_i \mathcal{P}_j=\delta_{ij}\mathcal{P}_i$, $\sum_{j\geq 0}\mathcal{P}_j=\ident$, $\mathcal{P}_0\pi=\pi$, and $\lvert\textrm{Re}\,\lambda_j\rvert < 1$. 
Using the fact that $\mathcal{M}$ leaves the trace invariant, one can show that $\bbra{1}$ is a left-eigenvector of $\mathcal{M}$ with eigenvalue~1. Assuming a unique steady state, it follows that
\[
    \mathcal{P}_0 = \kketbbra{\pi}{1}\,.
\]
Thus, observing for $\ell > 1$:
\[
\mathcal{M}^{\ell-1}= \mathcal{P}_0 + \sum_{j\geq 1} \lambda_j^{\ell-1} \mathcal{P}_j\,,
\]
and,
\begin{align*}
p_{y\gets x}(\ell) &= \frac{1}{p_x} \bbraopkket{1}{\mathcal{M}_y \kketbbra{\pi}{1} \mathcal{M}_x}{\pi} + \frac{1}{p_x}\sum_{j\geq 1} \lambda_j^{\ell-1} \bbraopkket{1}{\mathcal{M}_y \mathcal{P}_j \mathcal{M}_x}{\pi} \\
&= p_y + \frac{1}{p_x} \sum_{j\geq 1} \lambda_j^{\ell-1} \bbraopkket{1}{\mathcal{M}_y \mathcal{P}_j \mathcal{M}_x}{\pi} \\
p_{y\gets x}(1) &= \frac{1}{p_x} \bbraopkket{1}{\mathcal{M}_y \mathcal{M}_x}{\pi} \\
&= p_y + \frac{1}{p_x} \sum_{j\geq 1} \bbraopkket{1}{\mathcal{M}_y \mathcal{P}_j \mathcal{M}_x}{\pi}
\end{align*}
where the $\ell=1$ case is needed to address the indeterminate form that arises for $\lambda=0$ (i.e., $\mathcal{M}$ can have a nontrivial kernel). The last equality uses the decomposition $\sum_{j\geq 0}\mathcal{P}_j=\ident$.

The $\Psi$ matrix is then,
\begin{align*}
    \Psi_{yx} &= \sum_{\ell=1}^{\numMeas-1} \left( 1-\frac{\ell}{\numMeas} \right)\left[ p_{y\gets  x}(\ell)-p_y \right] \\
    &= \frac{1}{p_x} \sum_{j\geq 1} \left[ \left( 1-\frac{1}{\numMeas}\right) + \sum_{\ell=2}^{\numMeas-1} \left( 1-\frac{\ell}{\numMeas}\right) \lambda_j^{\ell-1} \right] \bbraopkket{1}{\mathcal{M}_y \mathcal{P}_j \mathcal{M}_x}{\pi} \\
    &= \frac{1}{p_x} \sum_{j\geq 1} \frac{1}{1-\lambda_j} \left[ 1 - \frac{1}{\numMeas} \frac{1-\lambda_j^{\numMeas}}{1-\lambda_j} \right] \bbraopkket{1}{\mathcal{M}_y \mathcal{P}_j \mathcal{M}_x}{\pi} \\
    &\overset{\numMeas\gg1}{\approx} \frac{1}{p_x} \bbra{1}{\mathcal{M}_y \left( \sum_{j\geq 1} \frac{1}{1-\lambda_j}\mathcal{P}_j \right) \mathcal{M}_x}\kket{\pi}
\end{align*}
using $\lvert\lambda_j\rvert < 1$ for $j\geq 1$ because these correspond to transient states.  
Denote the \keyword{Drazin inverse} of superoperator $\mathcal{A}$ as $\mathcal{A}^\draz$, so
\begin{equation}
    (\ident-\mathcal{M})^\draz = \sum_{j\geq 1} \frac{1}{1-\lambda_j}\mathcal{P}_j\,,
\end{equation}
observing that $\mathcal{P}_0$ projects to $\ker(\ident-\mathcal{M})$ and $\lambda_j\neq 1$ for $j\geq0$. Finally,
\begin{equation}\label{eq:app_PsiMat_quantJmps}
    \Psi_{yx} \overset{\numMeas\gg1}{\approx} \frac{1}{p_x} \bbraopkket{1}{\mathcal{M}_y (\ident-\mathcal{M})^\draz \mathcal{M}_x}{\pi}\,.
\end{equation}

\subsection{Sequence of measurements}\label{app:seq_measure}

As described in Ref.~\cite{smiga_stochastic_2023}, the calculations for the Fisher information contained in the empirical distribution of individual measurements in a stochastic process extends naturally to the distribution of finite sequences of measurements. 

Namely, the Fisher information for a sequence of $\seqLen$ is given by (for large-$\numMeas$)
\begin{equation}\label{eq:app_FI_emp_seq}
    F_\theta^{\textrm{(emp,}\seqLen\textrm{)}} = \numMeas \left( \partial_\theta \boldsymbol{p} \right)^\trans \left( \mathbb{P} + \Psi\mathbb{P} + \mathbb{P}\Psi^\trans \right)^{-1} \left( \partial_\theta \boldsymbol{p} \right)\,,
\end{equation}
which, at first, appears the same as Eq.~\eqref{eq:FI_emp_discrete}, except the vectors and matrices are now indexed by sequences of measurements instead of a single measurement. That is, $\boldsymbol{p}=\{\probof{\boldsymbol{x}}\}_{\boldsymbol{x}}$ over $\boldsymbol{x}=\{x_1,\ldots,x_\seqLen\}$ varying over all sequences of measurements, and $\mathbb{P}=\textrm{diag}\,\boldsymbol{p}$. Information about the correlations is then contained in the matrix,
\begin{equation}\label{eq:app_PsiMat_seq}
    \Psi_{\boldsymbol{y}\boldsymbol{x}} = \sum_{\ell=1}^{\numMeas-\seqLen} \left( 1-\frac{\ell}{\numMeas-\seqLen+1} \right)\left[ p_{\boldsymbol{y}\gets \boldsymbol{x}}(\ell)-p_{\boldsymbol{y}} \right]\,,
\end{equation}
for
\begin{equation}\label{eq:app_cond_observation_seq}
    p_{\boldsymbol{y}\gets \boldsymbol{x}}(\ell) = \probgiven{X_{i+\ell:i+\ell+\seqLen}=\boldsymbol{y}}{X_{i:i+\seqLen}=\boldsymbol{x}}\,, 
\end{equation}
using the shorthand $X_{a:b}=\{X_a,X_{a+1},\ldots,X_b\}$. The above equation describes the probability of measuring the sequence $\boldsymbol{y}$ after $\ell$ steps from measuring sequence $\boldsymbol{x}$ (counting from the start of the sequences).

Unlike the case of a single measurement, there is some influence of overlapping sequences when taking the sum in Eq.~\eqref{eq:app_PsiMat_seq}. For example, writing $\boldsymbol{x}=x_1x_2x_3$ and $\boldsymbol{y}=y_1y_2y_3$, then $p_{\boldsymbol{y}\gets \boldsymbol{x}}(2)$ vanishes unless $x_3=y_1$. Naturally, this introduces some inherent correlations between measurements, even if the individual measurements are independent, because overlapping sequences are correlated. For the most part, all of these considerations can be absorbed into the definition of $p_{\boldsymbol{y}\gets \boldsymbol{x}}(\ell)$, but they play a role in extending the calculation of $\Psi$ for quantum jumps described in Appendix~\ref{app:PsiMat_single}.

First, to simplify notation, let $\mathcal{M}_{\boldsymbol{x}}:=\mathcal{M}_{x_\seqLen}\cdots\mathcal{M}_{x_1}$.
Then, 
\begin{align}
    p_{\boldsymbol{y}} &= \bbraopkket{1}{\mathcal{M}_{\boldsymbol{y}}}{\pi}\,, \\
    p_{\boldsymbol{y}\gets \boldsymbol{x}}(\ell) &= \frac{1}{p_{\boldsymbol{x}}} \bbraopkket{1}{\mathcal{M}_{\boldsymbol{y}}\mathcal{M}^{\ell-\seqLen}\mathcal{M}_{\boldsymbol{x}}}{\pi}\,,
\end{align}
for $\ell\geq \seqLen$ and identifying $\mathcal{M}^{0}\to\ident$. When $1\leq\ell<\seqLen$,
\begin{equation}
    p_{\boldsymbol{y}\gets \boldsymbol{x}}(\ell) = \delta_{y_{\seqLen-\ell}x_\seqLen}\cdots\delta_{y_{1}x_{\ell+1}} \frac{1}{p_{\boldsymbol{x}}} \bbraopkket{1}{\mathcal{M}_{y_\seqLen}\cdots\mathcal{M}_{y_{\seqLen-\ell+1}} \mathcal{M}_{x_\seqLen}\cdots\mathcal{M}_{x_1}}{\pi}\,.
\end{equation}
To illustrate this function, consider the $\seqLen=2$ case so $\boldsymbol{x}=\{x_1,x_2\}$ and $\boldsymbol{y}=\{y_1,y_2\}$,
\[
p_{\boldsymbol{y}\gets \boldsymbol{x}}(\ell) = 
\begin{cases}
    \frac{\delta_{y_1x_2}}{p_{\boldsymbol{x}}} \bbraopkket{1}{\mathcal{M}_{y_2} \mathcal{M}_{x_2}\mathcal{M}_{x_1} }{\pi} & \ell = 1 \\
    \frac{1}{p_{\boldsymbol{x}}} \bbraopkket{1}{\mathcal{M}_{y_2}\mathcal{M}_{y_1}\mathcal{M}^{\ell-2} \mathcal{M}_{x_2}\mathcal{M}_{x_1} }{\pi} & \ell \geq 2
\end{cases}\,,
\]
and similarly for $\seqLen=3$,
\[
p_{\boldsymbol{y}\gets \boldsymbol{x}}(\ell) = 
\begin{cases}
    \frac{\delta_{y_2x_3}\delta_{y_1x_2}}{p_{\boldsymbol{x}}} \bbraopkket{1}{\mathcal{M}_{y_3} \mathcal{M}_{x_3}\mathcal{M}_{x_2}\mathcal{M}_{x_1} }{\pi} & \ell = 1 \\
    \frac{\delta_{y_1x_3}}{p_{\boldsymbol{x}}} \bbraopkket{1}{\mathcal{M}_{y_3}\mathcal{M}_{y_2} \mathcal{M}_{x_3}\mathcal{M}_{x_2}\mathcal{M}_{x_1} }{\pi} & \ell = 2 \\
    \frac{1}{p_{\boldsymbol{x}}} \bbraopkket{1}{\mathcal{M}_{y_3}\mathcal{M}_{y_2}\mathcal{M}_{y_1}\mathcal{M}^{\ell-3} \mathcal{M}_{x_3}\mathcal{M}_{x_2}\mathcal{M}_{x_1} }{\pi} & \ell \geq 3
\end{cases}\,.
\]

To simplify the calculation of $\Psi$, one can decompose the superoperator
\begin{equation}
    \mathcal{M} = \mathcal{P}_0 + \sum_{j\geq 1} \lambda_j \mathcal{P}_j\,,
\end{equation}
where, as before, $\mathcal{P}_i \mathcal{P}_j=\delta_{ij}\mathcal{P}_i$, $\sum_{j\geq 0}\mathcal{P}_j=\ident$, $\mathcal{P}_0\pi=\pi$, and $\lvert\textrm{Re}\,\lambda_j\rvert < 1$. It follows that $\mathcal{P}_0 = \kketbbra{\pi}{1}$, assuming there is a unique steady state.
Focusing on the part of the summation in Eq.~\eqref{eq:app_PsiMat_seq} without overlap ($\ell\geq\seqLen$),
\begin{align*}
    p_{\boldsymbol{y}\gets \boldsymbol{x}}(\ell)-p_{\boldsymbol{y}} &= \frac{1}{p_{\boldsymbol{x}}} \bbraopkket{1}{\mathcal{M}_{\boldsymbol{y}}\mathcal{M}^{\ell-\seqLen}\mathcal{M}_{\boldsymbol{x}}}{\pi}-p_{\boldsymbol{y}} \\
    &= \frac{1}{p_{\boldsymbol{x}}} \left[ \bbraopkket{1}{\mathcal{M}_{\boldsymbol{y}} \kketbbra{\pi}{1} \mathcal{M}_{\boldsymbol{x}}}{\pi} + \sum_{j\geq 1} \lambda_j^{\ell-\seqLen} \bbraopkket{1}{\mathcal{M}_{\boldsymbol{y}}\mathcal{P}_j\mathcal{M}_{\boldsymbol{x}}}{\pi} \right]-p_{\boldsymbol{y}} \\
    &= \frac{1}{p_{\boldsymbol{x}}} \sum_{j\geq 1} \lambda_j^{\ell-\seqLen} \bbraopkket{1}{\mathcal{M}_{\boldsymbol{y}}\mathcal{P}_j\mathcal{M}_{\boldsymbol{x}}}{\pi}\,.
\end{align*}
Then,
\begin{align*}
    \sum_{\ell=\seqLen}^{\numMeas-\seqLen} \left( 1-\frac{\ell}{\numMeas-\seqLen+1} \right) \left[ p_{\boldsymbol{y}\gets \boldsymbol{x}}(\ell)-p_{\boldsymbol{y}} \right] &= \frac{1}{p_{\boldsymbol{x}}} \sum_{j\geq 1} \sum_{\ell=\seqLen}^{\numMeas-\seqLen} \left( 1-\frac{\ell}{\numMeas-\seqLen+1} \right) \lambda_j^{\ell-\seqLen} \bbraopkket{1}{\mathcal{M}_{\boldsymbol{y}}\mathcal{P}_j\mathcal{M}_{\boldsymbol{x}}}{\pi} \\
    &= \frac{1}{p_{\boldsymbol{x}}} \sum_{j\geq 1} \left[ \frac{1}{1-\lambda_j} - \frac{(1-\lambda_j)(\seqLen-1) + 1-\lambda_j^{\numMeas-2(\seqLen-1)}}{(\numMeas-\seqLen+1)(1-\lambda_j)^2} \right] \bbraopkket{1}{\mathcal{M}_{\boldsymbol{y}}\mathcal{P}_j\mathcal{M}_{\boldsymbol{x}}}{\pi} \\
    &\overset{\numMeas\gg\seqLen}{\approx} \frac{1}{p_{\boldsymbol{x}}} \bbra{1}{\mathcal{M}_{\boldsymbol{y}}\left[\sum_{j\geq 1} \frac{\mathcal{P}_j}{1-\lambda_j} \right]\mathcal{M}_{\boldsymbol{x}}}\kket{\pi} \\
    &= \frac{1}{p_{\boldsymbol{x}}} \bbra{1}{\mathcal{M}_{\boldsymbol{y}}\left(1-\mathcal{M}\right)^{+}\mathcal{M}_{\boldsymbol{x}}}\kket{\pi}\,.
\end{align*}
The terms in the sum that include overlap are
\[
    \sum_{\ell=1}^{\seqLen-1} \left( 1-\frac{\ell}{\numMeas-\seqLen+1} \right) \left[ p_{\boldsymbol{y}\gets \boldsymbol{x}}(\ell)-p_{\boldsymbol{y}} \right] \overset{\numMeas\gg\seqLen}{\approx} \sum_{\ell=1}^{\seqLen-1} \left[ p_{\boldsymbol{y}\gets \boldsymbol{x}}(\ell)-p_{\boldsymbol{y}} \right]\,.
\]
Together,
\begin{equation}\label{eq:app_PsiMat_quantJmps_seq}
    \Psi_{\boldsymbol{y}\boldsymbol{x}} \overset{\numMeas\gg\seqLen}{\approx} \sum_{\ell=1}^{\seqLen-1} \left[ p_{\boldsymbol{y}\gets \boldsymbol{x}}(\ell)-p_{\boldsymbol{y}} \right] + \frac{1}{p_{\boldsymbol{x}}} \bbra{1}{\mathcal{M}_{\boldsymbol{y}}\left(1-\mathcal{M}\right)^{+}\mathcal{M}_{\boldsymbol{x}}}\kket{\pi}\,.
\end{equation}
Observe that the $\Psi$ matrix for a sequence of quantum jump measurements is very similar to the one previously derived [Eq.~\eqref{eq:app_PsiMat_quantJmps}], except for the addition of a finite sum to capture correlations between overlapping sequences.

\section{Product of non-zero eigenvalues}\label{app:ident_probdet}
In this section, we will show the following product of non-zero eigenvalues is given by
\begin{equation}
    \pdet\left( \mathbb{P}-\boldsymbol{p}\boldsymbol{p}^\trans \right) = d \prod_{i=1}^d p_i\,,
\end{equation}
where $\boldsymbol{p}=\{p_i\}_{i=1}^d$ (all non-zero), $\sum_{i=1}^d p_i = 1$, and $\mathbb{P}=\textrm{diag}\,\boldsymbol{p}$.

\begin{proof}
    First, consider the nullspace for $(\mathbb{P}-\boldsymbol{p}\boldsymbol{p}^\trans)$. From the condition $\sum_{i=1}^d p_i = 1$, it follows that the nullspace contains
    \begin{equation}
        \onevec = \{ 1,1,\ldots,1\}\,.
    \end{equation}
    Then, 
    \begin{equation}
        \pdet\left( \mathbb{P}-\boldsymbol{p}\boldsymbol{p}^\trans \right) = \det\left( \mathbb{P}-\boldsymbol{p}\boldsymbol{p}^\trans + \frac{1}{d}\onevec\onevec^\trans \right)\,,
    \end{equation}
    where the additional term $\frac{1}{d}\onevec\onevec^\trans$ ``lifts'' the nullspace to the 1-eigenspace without affecting the other eigenvalues. 
    Note that if the nullspace is multi-dimensional, additional terms would need to be added. This case would result in a determinant of zero which cannot occur for the product of finitely many positive numbers, so this would be a clear failure point of the proof. For the time being, we will assume that the nullspace is 1-dimensional.

    For the remainder of the proof follows from two identities. Let $A$ be an invertible matrix and $\boldsymbol{u}$ and $\boldsymbol{v}$ vectors. 
    First, the \keyword{Sherman-Morrison formula} states that if $1+\boldsymbol{v}^\trans A^{-1} \boldsymbol{u} \neq 0$, then
    \begin{equation}
        (A+\boldsymbol{u}\boldsymbol{v}^{\rm T})^{-1}  =A^{-1} - \frac{A^{-1}\boldsymbol{u}\boldsymbol{v}^{\rm T}A^{-1}}{1+\boldsymbol{v}^{\rm T}A^{-1}\boldsymbol{u}}\,.
    \end{equation}
    Second, the \keyword{matrix determinant lemma} is
    \begin{equation}
        \det\left( A+\boldsymbol{u}\boldsymbol{v}^{\rm T}\right) = \left(1+\boldsymbol{v}^{\rm T}A^{-1}\boldsymbol{u}\right) \det{A}\,.
    \end{equation}

    Now,
    \[
        \det\left( \mathbb{P}-\boldsymbol{p}\boldsymbol{p}^\trans + \frac{1}{d}\onevec\onevec^\trans \right) = \left[ 1-\boldsymbol{p}^\trans\left( \mathbb{P}+\frac{1}{d}\onevec\onevec^\trans \right)^{-1}\boldsymbol{p} \right] \det\left(\mathbb{P}+\frac{1}{d}\onevec\onevec^\trans\right)\,.
    \]
    Then
    \[
        \det\left(\mathbb{P}+\frac{1}{d}\onevec\onevec^\trans\right) = \left( 1+\frac{1}{d}\onevec^\trans\boldsymbol{b} \right) \det\mathbb{P}\,,
    \]
    where $\boldsymbol{b}=\mathbb{P}^{-1}\onevec=\{1/p_i\}_{i=1}^d$. It follows immediately that $\boldsymbol{b}\cdot\boldsymbol{p}=d$.
    Also,
    \begin{align*}
        \left( \mathbb{P}+\frac{1}{d}\onevec\onevec^\trans \right)^{-1} &= \mathbb{P}^{-1} - \frac{\mathbb{P}^{-1}\onevec\onevec^\trans \mathbb{P}^{-1}/d}{1+\onevec^\trans \mathbb{P}^{-1}\onevec/d} \\
        &= \mathbb{P}^{-1} - \frac{\boldsymbol{b}\boldsymbol{b}^\trans/d}{1+\onevec^\trans\boldsymbol{b}/d}\,,
    \end{align*}
    so
    \[
        \boldsymbol{p}^\trans\left( \mathbb{P}+\frac{1}{d}\onevec\onevec^\trans \right)^{-1}\boldsymbol{p} = 1-\frac{d}{1+\onevec^\trans\boldsymbol{b}/d}\,.
    \]
    Together,
    \[
        \det\left( \mathbb{P}-\boldsymbol{p}\boldsymbol{p}^\trans + \frac{1}{d}\onevec\onevec^\trans \right) = \frac{d}{1+\onevec^\trans\boldsymbol{b}/d} \left( 1+\frac{1}{d}\onevec^\trans\boldsymbol{b} \right) \det\mathbb{P} = d \prod_{i=1}^d p_i\,,
    \]
    because $\det\mathbb{P}=\prod_{i=1}^d p_i$.
\end{proof}

\section{Constraints on distribution of finite sequences}\label{app:seq_constraints}
When considering the information contained in the empirical distribution of finite sequences of measurements, there are additional constraints on the support of the empirical distribution. This arises from the fact the number of times going into a sequence of measurement must be the same as the number of times going out of that sequence of measurement. 

As a simple example, consider an $\seqLen=2$ sequence of binary measurements in the alphabet $\mathbb{M}=\{0,1\}$. A given sequence of measurements,
\[
    11\overline{10}000\underline{01}1\overline{10}00\underline{01}11111\,.
\]
As highlighted by the over/underlines, occurrences of $01$ happen the same number of times as $10$. No matter how long the sequence or what the underlying dynamics, the number of times $01$ appears differs by the number of times $10$ appears by \emph{at most} one. 
In contrast if one were to ``reset'' the sequence between measuring pairs, there are no hard limits as to the relative number of $01$ and $10$ sequences; though they would converge to the same value in the large-$\numMeas$ limit. 

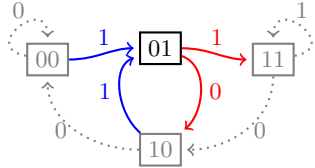
\begin{figure}[ht]
    \centering
    \begin{tikzpicture}[x=15mm,y=15mm,shorten >=2pt, thick, 
                        focus node/.style={rectangle,draw}, 
                        side node/.style={rectangle,draw,gray}
                        ]
        \node[focus node] (01) at (0,0.1)  {01};
        \node[side node]  (00) at (-1,0) {00};
        \node[side node]  (10) at (0,-.8) {10};
        \node[side node]  (11) at (1,0)  {11};
    
        \path [->,out=0,in=180,blue] (00) edge node [above] {1} (01);
        \path [->,out=140,in=200,blue] (10) edge node [left] {1} (01);
        \path [->,out=0,in=180,red] (01) edge node [above] {1} (11);
        \path [->,out=340,in=40,red] (01) edge node [right] {0} (10);
        
        \begin{scope}[dotted,gray] 
            \path [->] (00) edge [out=170,in=90,looseness=5] node [above] {0} (00);
            \path [->] (11) edge [out=10,in=90,looseness=5] node [above] {1} (11);
            \path [->] (11) edge [out=270,in=0] node [right] {0} (10);
            \path [->] (10) edge [out=180,in=270] node [left] {0} (00);
        \end{scope}
    \end{tikzpicture}
    
    \caption{Graph of observing sequence that emphasizes going ``into'' and ``out of'' sequences of data from a $d=2$-alphabet. This diagram is styled to highlight a constraint that appears on $\seqLen=3$ empirical distributions, namely for $\vec{x}^{\seqLen-1}=01$. Comparing to Eq.~\eqref{eq:constraint_inout}, the nodes are labeled by $\vec{x}^{\seqLen-1}$ and the edges are labeled by $z$. The total occurrences of a processes passing edges into $01$ (blue) must equal the number of occurrences exiting $01$ (red).}
    \label{fig:constraint_vis}
\end{figure}

The general constraint arises from equating how many times going into and out of each $(\seqLen-1)$-sequence. Denoting concatenation of a measurement $x$ and $y$ with $xy$ and a $k$-sequence with $\vec{x}^k=x_1x_2\ldots x_k$. Then this constraint can be understood as, for all $\vec{x}^{\seqLen-1}$,
\begin{equation}\label{eq:constraint_inout}
    \sum_z q_{z\vec{x}^{\seqLen-1}} = 
    \sum_{z'} q_{\vec{x}^{\seqLen-1}z'}+\varepsilon_\numMeas\,,
\end{equation}
where $q_{\vec{x}^{\seqLen}}$ is the empirical distribution of length-$\seqLen$ sequences derived from any $\numMeas$ sequence of data. 
If the alphabet consists of $d$ entries, this corresponds to $d^{\seqLen-1}$ constraints.
The error $\varepsilon_\numMeas\in\{0,\pm \frac{1}{\numMeas-\seqLen+1}\}$ can be non-zero when the measurements start with $\vec{x}^{\seqLen-1}$ but end with $\vec{y}^{\seqLen-1}\neq\vec{x}^{\seqLen-1}$; in which there would be one $\frac{+1}{\numMeas-\seqLen+1}$ and one $\frac{-1}{\numMeas-\seqLen+1}$ error; likewise, knowing the first and last $(\seqLen-1)$ measurements will fully fix this error. 
Given the strict bound on this error, we will neglect it for the remainder of this appendix; it does not meaningfully affect the support of the empirical distribution.
A visualization of this constraint in the $\seqLen=3$ and $d=2$ case with emphasis on $\vec{x}^{\seqLen-1}=01$ is shown in Fig.~\ref{fig:constraint_vis}. 
For notational simplicity, denote $\vec{x}^{\seqLen-1}=\vec{x}$ while $\boldsymbol{y}$ denotes the complete $\seqLen$-sequence. 

There is also a normalization constraint, as before,
\begin{equation}\label{eq:constraint_norm}
    \sum_{\boldsymbol{z}} q_{\boldsymbol{z}} = 1\,.
\end{equation}
However, this with the constraints from Eq.~\eqref{eq:constraint_inout} have a redundancy. To see this, write
\begin{align*}
    I_{\vec{x}} &= \sum_z q_{z\vec{x}} \\
    O_{\vec{x}} &= \sum_z q_{\vec{x}z} \,,
\end{align*}
where Eq.~\eqref{eq:constraint_inout} becomes $I_{\vec{x}}=O_{\vec{x}}$.
For a fixed sequence $\vec{y}$, we would like to show that the normalization constraint, Eq.~\eqref{eq:constraint_norm}, and all the constraints in Eq.~\eqref{eq:constraint_inout} except for $I_{\vec{y}}=O_{\vec{y}}$ can be used to derive the latter constraint. 
Equation~\eqref{eq:constraint_norm} implies $\sum_{\vec{x}} I_{\vec{x}} = \sum_{\vec{x}} O_{\vec{x}} =1$ regardless of the constraints in Eq.~\eqref{eq:constraint_inout}. Using this, one can write $\sum_{\vec{x}\setminus \vec{y}} I_{\vec{x}}=1-I_{\vec{y}}$ and $\sum_{\vec{x}\setminus \vec{y}} O_{\vec{x}}=1-O_{\vec{y}}$, where the sums exclude $\vec{y}$.
Then adding up all the constraints in Eq.~\eqref{eq:constraint_inout} except for $\vec{x}=\vec{y}$,
\begin{align*}
    \sum_{\vec{x}\setminus \vec{y}} I_{\vec{x}} &= \sum_{\vec{x}\setminus \vec{y}} O_{\vec{x}} \\
    1-I_{\vec{y}} &= 1-O_{\vec{y}} \\
    I_{\vec{y}} &= O_{\vec{y}}\,,
\end{align*}
the final constraint.
In other words, knowing that there is the normalization condition for empirical distributions and knowing the ``flow'' in/out of all but one sequence $\vec{y}$ is enough to know the flow in/out of $\vec{y}$. 
This leads to $d^{\seqLen-1}$ constraints in total on the space of $d^\seqLen$-dimensional empirical distributions.

From the nominal $d^\seqLen$-dimensional space containing the empirical distribution, the constraints would restrict to a hypersurface with codimension $d^{\seqLen-1}$. However, this neglects the possibility of an error term $\varepsilon_\numMeas\in\{0,\pm \frac{1}{\numMeas-\seqLen+1}\}$. When included, the range of possible empirical distributions extends slightly away from this hypersurface, though to a region still much smaller than the full $d^\seqLen$-dimensional space and whose width shrinks as $\mathcal{O}\left( \numMeas^{-1} \right)$. In general, the error introduced by approximating the empirical distribution to a nearby point on the hypersurface will not meaningfully affect the calculations.

Using the constraints, one can ``throw out'' some of the information contained in the empirical distribution $q_{\boldsymbol{x}}$. 
Any reduced set of information must be able to recreate the full empirical distribution. Consider removing all elements in the empirical distribution that start with $y$, leaving $d^\seqLen-d^{\seqLen-1}$ elements. One can reconstructing the lost elements using the constraints. 
First, consider all $\vec{x}$ that do not start with $y$. Using Eq.~\eqref{eq:constraint_inout}, ignoring the error\footnote{The potential presence of an error $\varepsilon_\numMeas$ can lead to some ambiguity after omitting values, but this error is strictly bounded, so it will not affect the overall statistics.},
\[
    q_{y\vec{x}} = \sum_{z'} q_{\vec{x}z'} - \sum_{z\setminus y} q_{z\vec{x}}\,,
\]
defines the $q_{y\vec{x}}$ entries using only entries that do not start with $y$. 
Then, for $\vec{x}=y\vec{w}^{\seqLen-2}$ where $\vec{w}^{\seqLen-2}$ does not start with $y$, 
\[
    q_{yy\vec{w}^{\seqLen-2}} = \sum_{z'} q_{y\vec{w}^{\seqLen-2}z'} - \sum_{z\setminus y} q_{zy\vec{w}^{\seqLen-2}}\,.
\]
The right-hand side of this equation depends only on known or derived elements. 
Repeating this recursively can be used to calculate every entry except $q_{yy\cdots y}$. This final entry can be calculated with Eq.~\eqref{eq:constraint_norm},
\[
    q_{yy\cdots y}=1-\sum_{\boldsymbol{x}\setminus yy\cdots y}q_{\boldsymbol{x}}\,.
\]
It likewise follows that one could omit all elements that end with $y$, for any fixed $y$. 
One can think of these omissions as a type of lossless compression or filtering. Further, the existence of a filter leaving a $(d^\seqLen-d^{\seqLen-1})$-dimensional object confirms that the $d^{\seqLen-1}$ constraints are not redundant. 

\end{document}